\documentclass[]{aastex631}

\graphicspath{{./}{figures/}}

\usepackage{amsmath}
\usepackage{hyperref}
\defcitealias{2013A&A...557A..52P}{PCXI}

\newcommand{\Planck}{{\em Planck}}

\begin{document}

\title{Probing Hot Gas Components of Circumgalactic Medium in Cosmological Simulations with the Thermal Sunyaev-Zel'dovich Effect}

\author[0000-0002-4274-9373]{Junhan Kim}
\email{junhank@caltech.edu}
\affiliation{California Institute of Technology, 1200 E. California Blvd., MC 367-17, Pasadena, CA 91125, USA}

\author[0000-0002-1098-7174]{Sunil Golwala}
\affiliation{California Institute of Technology, 1200 E. California Blvd., MC 367-17, Pasadena, CA 91125, USA}

\author{James G. Bartlett}
\affiliation{Université de Paris, CNRS, Astroparticule et Cosmologie, F-75013 Paris, France}
\affiliation{Jet Propulsion Laboratory, California Institute of Technology, 4800 Oak Grove Drive, Pasadena, CA 91109, USA}

\author[0000-0002-4200-9965]{Stefania Amodeo}
\affiliation{Universit\'e de Strasbourg, CNRS, Observatoire astronomique de Strasbourg, UMR 7550, F-67000 Strasbourg, France}
\affiliation{Department of Astronomy, Cornell University, Ithaca, NY 14853, USA}

\author[0000-0001-5846-0411]{Nicholas Battaglia}
\affiliation{Department of Astronomy, Cornell University, Ithaca, NY 14853, USA}

\author[0000-0001-5501-6008]{Andrew J. Benson}
\affiliation{Carnegie Observatories, 813 Santa Barbara Street, Pasadena, CA 91101, USA}

\author[0000-0002-9539-0835]{J. Colin Hill}
\affiliation{Department of Physics, Columbia University, New York, NY 10027, USA}
\affiliation{Center for Computational Astrophysics, Flatiron Institute, 162 Fifth Avenue, New York, NY 10010, USA}

\author[0000-0003-3729-1684]{Philip F. Hopkins}
\affiliation{California Institute of Technology, TAPIR, MC 350-17, Pasadena, CA 91125, USA}

\author[0000-0002-3817-8133]{Cameron B. Hummels}
\affiliation{California Institute of Technology, TAPIR, MC 350-17, Pasadena, CA 91125, USA}

\author[0000-0003-1593-1505]{Emily Moser}
\affiliation{Department of Astronomy, Cornell University, Ithaca, NY 14853, USA}

\author[0000-0003-1053-3081]{Matthew E. Orr}
\affiliation{California Institute of Technology, TAPIR, MC 350-17, Pasadena, CA 91125, USA}
\affiliation{Department of Physics and Astronomy, Rutgers University, 136 Frelinghuysen Road, Piscataway, NJ 08854, USA}
\affiliation{Center for Computational Astrophysics, Flatiron Institute, 162 Fifth Avenue, New York, NY 10010, USA}

\begin{abstract}
The thermal Sunyaev-Zel'dovich (tSZ) effect is a powerful tool with the potential for constraining directly the properties of the hot gas that dominates dark matter halos because it measures pressure and thus thermal energy density. Studying this hot component of the circumgalactic medium (CGM) is important because it is strongly impacted by star-formation and active galactic nuclei (AGN) activity in galaxies, participating in the feedback loop that regulates star and black hole mass growth in galaxies.

We study the tSZ effect across a wide halo mass range using three cosmological hydrodynamical simulations: Illustris-TNG, EAGLE, and FIRE-2. Specifically, we present the scaling relation between tSZ signal and halo mass and radial profiles of gas density, temperature, and pressure for all three simulations. The analysis includes comparisons to \Planck\ tSZ observations and to the thermal pressure profile inferred from the Atacama Cosmology Telescope (ACT) measurements. We compare these tSZ data to the simulations to interpret the measurements in terms of feedback and accretion processes in the CGM. We also identify as-yet unobserved potential signatures of these processes that may be visible in future measurements, which will have the capability of measuring tSZ signals to even lower masses. We also perform internal comparisons between runs with different physical assumptions. We conclude: (1) there is strong evidence for the impact of feedback at $R_{500}$ but that this impact decreases by $5R_{500}$, and (2) the thermodynamic profiles of the CGM are highly dependent on the implemented model, such as cosmic-ray or AGN feedback prescriptions.
\end{abstract}

\section{Introduction} \label{sec:intro}

With the development of observational tools and techniques, we are gaining a better understanding of the gas surrounding galaxies, known as the circumgalactic medium (CGM; see \citealt{2017ARA&A..55..389T} for a recent review). The CGM and the intracluster medium (ICM) in galaxy clusters hold the majority of baryons as a result of multiple processes that distribute (and redistribute) them as a diffuse gas within dark matter halos over cosmic time. The baryons in the CGM, both those suspended there due to accretion-shock heating and those expelled from galaxies and their interstellar medium (ISM), are believed to constitute the ``missing baryons'' \citep[e.g.,][]{1992MNRAS.258P..14P, 1998ApJ...503..518F, 2007ARA&A..45..221B}. Therefore, the thermodynamic properties of the CGM afford useful information for understanding the physical processes governing galaxy formation and evolution. 

Until recently, CGM studies have mostly relied on ultraviolet (UV) and optical absorption line observations, probing warm and cold components of the CGM \citep[e.g.,][]{2017MNRAS.465.2966S, 2018MNRAS.477..450N}. The Sunyaev-Zel'dovich (SZ) effect (\citealt{1970Ap&SS...7....3S, 1972CoASP...4..173S, 1980MNRAS.190..413S}; see \citealt{2019SSRv..215...17M} for a recent review) and soft X-ray emission have become powerful new tools to study the multiphase CGM's hot gas component, whose temperature exceeds $\sim10^6$~K. These tools provide a calorimetric view of the ionized gas that complements absorption line observations.

The thermal SZ (tSZ) effect is a spectral distortion of the cosmic microwave background (CMB) radiation caused by inverse Compton scattering of CMB photons by free electrons. The signal is proportional to the electron thermal pressure of the halo gas, integrated along the line-of-sight. Individual tSZ detections have been limited to massive clusters because of sensitivity and angular resolution, but stacked measurements of galaxy samples (e.g., \citealt{2013A&A...557A..52P}, \citetalias{2013A&A...557A..52P} hereafter; \citealt{2015ApJ...808..151G}) presented the mass scaling relation down to halo masses below $\sim 10^{13}\ \textrm{M}_\odot$. Notably, the \citetalias{2013A&A...557A..52P} results suggested that the integrated tSZ flux within a radius $R_{500}$\footnote{The radius within which the mean overdensity is 500 times the critical density at a given redshift.} followed a self-similar dependence on halo mass over more than two orders of magnitude in mass. The mass scaling derived from a self-similar relation between the gas pressure profile and halo mass \citep{1986MNRAS.222..323K, 1991ApJ...383..104K} implied that the total thermal energy was governed primarily by the gravitational potential. This self-similarity suggested by \citetalias{2013A&A...557A..52P} appears counter-intuitive because the non-gravitational feedback mechanisms that regulate star formation and black hole growth in galaxies are expected to inject energy into and heat the CGM \citep[e.g.,][]{2009MNRAS.399.1773C, 2010MNRAS.406.2325O, 2010MNRAS.402.1536S}. Specifically, stellar winds and supernovae (`stellar feedback' due to star formation) and active galactic nuclei (AGN; `AGN feedback') are expected to eject gas from the disks of galaxies into the CGM or out past it into the intergalactic medium (IGM) in a manner that depends on various halo properties such as halo mass, physics of star formation, and black hole mass accretion efficiency. However, the \citetalias{2013A&A...557A..52P} result was driven by the extrapolation of results measured at large radii ($5R_{500}$) to $R_{500}$, assuming a fixed pressure profile model of \cite{2010A&A...517A..92A} based on \cite{2007ApJ...668....1N} profile. Also, \cite{2018PhRvD..97h3501H} pointed out that neglecting the signal from nearby halos (`two-halo term') in the modeling could also have biased the inference toward a self-similar model. Therefore, a careful comparison between the tSZ observation and models, with an emphasis on the distribution of the hot gas and its properties, is required.

Observations using tSZ have already proven their worth in probing feedback \citep[e.g.,][]{verdier2016, 2016MNRAS.458.1478C, 2018ApJ...865..109S, 2021ApJ...913...88M}. \cite{2017JCAP...11..040B} furthermore demonstrated that combined measurements of the tSZ and the kinetic SZ (kSZ) effects provide constraints on thermodynamic processes affecting the gas. Recently, \cite{2021PhRvD.103f3513S} and \cite{2021PhRvD.103f3514A} cross-correlated the Atacama Cosmology Telescope (ACT) CMB survey maps \citep{2020JCAP...12..046N} with the Baryon Oscillation Spectroscopic Survey \citep[BOSS;][]{2014ApJS..211...17A} CMASS (``constant stellar mass'') galaxy catalog to obtain  stacked tSZ and kSZ measurements. Comparing the radial thermodynamic profiles from the measurements to the predictions from cosmological simulations, they found that the simulations tend to underpredict the gas pressure and density, notably around the halos' outer regions. Future millimeter-wave SZ instruments with improved angular resolution and sensitivity will provide data from halos down to galaxies and group scales \citep{2019BAAS...51c.124M}, enabling us to effectively probe feedback mechanisms in these lower mass halos \citep{2019BAAS...51c.297B}. Such instruments include next-generation CMB experiments \citep[e.g., CMB-S4, Simons Observatory;][]{2016arXiv161002743A, 2019arXiv190704473A, 2017arXiv170602464A, 2019JCAP...02..056A} and single large-aperture telescopes \citep[e.g. CSST, CMB-HD, AtLAST;][]{2018alas.confE..46G, 2019BAAS...51g...6S, 2019BAAS...51g..58K}.

While instrumentation and telescope development have enabled higher-sensitivity and higher-resolution observations, computational simulations have made significant progress as well. Current cosmological simulations have been successful in reproducing many observed galaxy properties, improving our understanding of galaxy formation and evolution \citep[see][for a recent review]{2020NatRP...2...42V}. However, these simulations often do not resolve the physical scales necessary to compute many astrophysical processes and instead resort to `sub-grid' prescriptions. The sub-grid prescriptions are coarse-level approximations of what physical mechanisms are occurring on finer resolution scales than are self-consistently modeled in the simulation. While these sub-grid models are calibrated against key observables, they are still a major source of uncertainty in computational models of galaxy evolution, so it behooves us to analyze multiple simulations with different sub-grid prescriptions. These physically motivated recipes give predictions for CGM properties that vary significantly depending on the models \citep[e.g.,][]{2012MNRAS.423.2991V, 2013MNRAS.430.1548H}. In particular, a comparison of tSZ simulations to data can distinguish which prescriptions are correct and thereby inform our understanding of the CGM and feedback mechanisms that affect it.

In this work, we study the tSZ effect and relevant gas properties using state-of-the-art simulations. These include two large-volume simulations, Illustris-TNG \citep[][The Next Generation; TNG hereafter]{2018MNRAS.475..624N, 2018MNRAS.475..648P, 2018MNRAS.475..676S, 2018MNRAS.477.1206N, 2018MNRAS.480.5113M} and Evolution and Assembly of GaLaxies and their Environments \citep[EAGLE,][]{2015MNRAS.446..521S, 2016A&C....15...72M, 2017arXiv170609899T}, and one zoom-in simulation, Feedback In Realistic Environments-2 \citep[FIRE-2,][]{2018MNRAS.480..800H, 2020MNRAS.496.1620O}. In Section~\ref{sec:sims_methods}, we describe the simulations and how we calculate the thermodynamic properties of the CGM and the tSZ signal. In Section~\ref{sec:tsz}, we compare the tSZ signal predicted by different models to the observations and perform internal comparisons between simulations. In particular, we (1) study scaling relations for integrated tSZ flux using TNG and EAGLE, separated by galaxy type, (2) compare EAGLE's predictions to the radial profiles inferred recently from ACT data, and (3) discuss radial profiles of Milky Way-sized ($\sim 10^{12} \textrm{M}_\odot$) halos at $z=0$ from TNG, EAGLE, and FIRE-2. For the latter, we include the recent FIRE-2 simulations that model cosmic-rays \citep{2020MNRAS.492.3465H, 2020MNRAS.496.4221J}. We then (4) discuss the effect of AGN feedback on the tSZ flux by comparing EAGLE simulations with and without the AGN feedback prescription. Finally, we summarize the implications of our results in Section~\ref{sec:conclusion} and discuss possible synergies of tSZ observations with other CGM probes in Section~\ref{sec:futurework}.

\section{Simulations and methods}\label{sec:sims_methods}

\subsection{Thermal SZ effect}

The tSZ effect is a distortion of the thermal frequency spectrum of the CMB caused by inverse Compton scattering of CMB photons with free electrons in any hot gas (including the CGM). The net effect is a shift of photons from lower to higher frequencies. The resultant spectrum is unique and has the same frequency dependence for all objects in the limit that the electrons are nonrelativistic.\footnote{Relativistic corrections cause the frequency dependence to depend on the electron temperature for temperatures approaching the electron rest energy ($T\gtrsim 10^8\,$K; e.g., \citealt{2012MNRAS.426..510C}), but the effect is negligible for typical CGM temperatures.} The amplitude of the distortion is proportional to the electron pressure integrated along the line-of-sight, as characterized by the Compton-$y$ parameter, 
\begin{equation}
    \begin{split}
        y &\equiv \frac{\sigma_T}{m_e c^2} \int P_e dl \\
        &= \frac{\sigma_T}{m_e c^2} \int n_e k_B T_e dl,
    \end{split}
\end{equation}
where $n_e$ is the electron number density, $\sigma_T$ is the Thomson cross-section, $k_B$ is the Boltzmann constant, and $T_e$ is the electron temperature; $m_e$ is the electron rest mass and $c$ is the speed of light. We describe the details of computing the tSZ signal using simulation dataset in Appendix~\ref{sec:appendix_tszcalc}. The literature, including \citetalias{2013A&A...557A..52P}, often quotes the parameter integrated within a sphere of a radius $R_{500}$:  
\begin{equation}
    \begin{split}
        Y_{R_{500}} {D^2_A (z)} &\equiv \int_{0}^{R_{500}} n_e \sigma_T \frac{k_B T_e}{m_e c^2} dV\\
        &= \int_{0}^{R_{500}} n_e \sigma_T \frac{k_B T_e}{m_e c^2} 4 \pi r^2 dr,
    \end{split}
\end{equation}
where $D_A (z)$ is the angular diameter distance to the given redshift. In order to put sources at different redshifts on the same footing (the equivalent of converting from flux to luminosity for more traditional measurements), we scale to $z=0$ and a fixed angular diameter distance of 500~Mpc,
\begin{equation}
    \label{eq:y_r500}
    \tilde{Y}_{R_{500}} = Y_{R_{500}} E^{-2/3}(z) \left[D_\textrm{A} (z) / 500 \textrm{Mpc}\right]^2,
\end{equation}
where $E(z)$ is the Hubble parameter $H(z)$ normalized by $H_0$, which is approximated by $E^2(z) = \Omega_\textrm{m} (1+z)^3 + \Omega_\Lambda$. We express $\tilde{Y}_{R_{500}}$ in units of arcmin$^2$.\footnote{Some authors do not scale to $D_A = 500$~Mpc, resulting in a quantity in Mpc$^2$ rather than arcmin$^2$.} Since the temperature is proportional to $M^{2/3}$ for a virialized halo, the tSZ relation is dependent only on $M$, i.e., $\sim M^{5/3}$, in the self-similar limit, in which gravity and thus mass determines all the halo parameters.

However, the spherically integrated $Y_{R_{500}}$ is not a direct observable due to the instrumental beam size and the contribution of all gas along the line-of-sight. Because of \Planck's several-arcmin beam, \citetalias{2013A&A...557A..52P} actually measured the signal within a cylindrical volume of projected radius $5\times R_{500}$ and then extrapolated to $Y_{R_{500}}$ assuming a universal electron pressure radial profile \citep{2010A&A...517A..92A}. In addition, the line-of-sight integration can be affected by the emission from nearby halos (`two-halo term'), thermal emission from dust in the host galaxies, the viewing angle, or other effects. The effect of the two-halo term on calculating the radial density and pressure profiles was discussed in several literatures \citep[e.g.,][]{2017MNRAS.467.2315V, 2018PhRvD..97h3501H, 2021ApJ...919....2M}, and it becomes non-negligible beyond a few times the virial radius. Following their approach, we include gas particles that do not belong to a central subhalo for the tSZ calculation in this work and discuss this issue in Appendix~\ref{sec:appendix_twohalo}.

\subsection{Simulations}

We use three simulation datasets, namely TNG, EAGLE, and FIRE-2, to compare tSZ signatures and gas radial profiles among galaxy samples. The simulations were tuned to reproduce various observed galaxy properties, such as the stellar mass function and galaxy size, employing different sub-grid prescriptions \citep{2015MNRAS.446..521S, 2018MNRAS.473.4077P, 2018MNRAS.480..800H}. Each of the simulations offers predictions for the distribution of the CGM in and around halos and for its thermodynamic properties that are highly dependent on its assumed model.

We mainly use publicly available TNG100-2\footnote{http://www.tng-project.org/} \citep{2019ComAC...6....2N} and EAGLE\footnote{http://icc.dur.ac.uk/Eagle/} Ref-L100N1504 in similar comoving volumes ($\sim$100$^3$~cMpc$^3$; cMpc: comoving Mpc). We focus our analysis on the present-day ($z=0$) snapshots from both simulations. To test the impact of AGN feedback, we also use lower-resolution, smaller-volume EAGLE simulations with and without feedback (Ref-L0050N0752 and NoAGN-L0050N0752), which use a comoving volume of 50$^3$~cMpc$^3$ with 752$^3$ smoothed particle hydrodynamics (SPH) and dark matter (DM) particles.  TNG and EAGLE provide halo (group) and subhalo (galaxy) catalogs identified using the friends-of-friends \citep[FoF;][]{1982ApJ...257..423H, 1985ApJ...292..371D} and \texttt{subfind} algorithms \citep{2001MNRAS.328..726S}. We only identify central `subhalos,' which are the equivalent of central galaxies, as galaxies for the analysis so that we can be assured that each galaxy's properties are driven by its own evolution rather than being dominated by a nearby larger galaxy. We do, however, include all nearby particles within a simulation volume to analyze gas properties, even if they are not bound to the specific halo/subhalo, because the CGM of a central galaxy and its satellites are not separate. We will describe the rationale for including these `two-halo' effects in detail in Appendix~\ref{sec:appendix_twohalo}.

We complement these recent state-of-the-art cosmological simulations with the `zoom-in' FIRE-2 simulation, which explores much smaller physical scales than the other two large-volume simulations and focuses on individual halos, because its finer physical resolution has already provided additional useful understanding of CGM properties \citep[e.g.,][]{2020arXiv200613976S}. In particular, in Section~\ref{subsec:radial}, we compare the gas radial profiles of the Milky Way-sized galaxies in FIRE-2 to average profiles of galaxies of similar mass from TNG and EAGLE. We assess the impact of cosmic-rays by also comparing to the recent FIRE-2 run with cosmic-ray treatment \citep{2020MNRAS.496.4221J}.

CGM properties in these simulations have been analyzed in several prior works. \cite{2018MNRAS.477..450N} analyzed TNG simulation data to study distribution of high ionization species O VI, O VII, and O VIII and their physical properties, which could be observed by UV and X-ray spectroscopy. \cite{2019MNRAS.485.3783D} studied the relation between the gas fraction and the halo mass of present-day $\sim\!L^{\star}$ galaxies in EAGLE. They showed that the central black hole mass and the halo gas fraction are strongly negatively correlated at a fixed halo mass and that the soft X-ray luminosity and the tSZ flux display similar correlations with the black hole mass and the star-formation rate (SFR). \cite{2020MNRAS.491.4462D} extended this work to TNG, finding that the scatter in CGM mass fraction is strongly correlated with the specific SFR (sSFR) in both EAGLE and TNG while the CGM fractions in low-mass halos are considerably higher in TNG than EAGLE. This result implies that tSZ observations of galaxy samples separated by their physical properties will give useful information about the halo gas and distinguish models.

\cite{2016MNRAS.463.4533V} studied the tSZ signal and the soft X-ray emission in the original FIRE \citep[FIRE-1;][]{2014MNRAS.445..581H} simulation to understand the effect of stellar feedback. (The FIRE-1 simulation did not implement AGN feedback.) They found that the tSZ flux (integrated only out to $R_{500}$ but including all the gas along the line-of-sight inside that radius in the simulation volume) of halos whose masses are below $10^{13}~\textrm{M}_\odot$ deviate from the \citetalias{2013A&A...557A..52P} self-similar scaling relation. They were able to explain this suppression in these low-mass objects by their reduced hot gas ($T > 10^{4}$~K) fraction, which is also observed to be redshift-independent, unlike the baryon and total gas fractions. They explained that the little dependence on the redshift was because high redshift halos contained more cold gas than hot gas, and they lost a smaller fraction of the baryons than low redshift halos, canceling out the effect on the hot gas fraction.


\section{Thermal SZ signal in Cosmological Simulations}\label{sec:tsz}

The tSZ signal offers a way to probe hot gas components around the halos by measuring the integrated electron pressure. It has been observed with several millimeter instruments and wide-field, high-sensitivity CMB surveys. The observational effort includes experiments such as \Planck\ \citep{2011A&A...536A..10P, 2013A&A...557A..52P}, ACT \citep{2016JLTP..184..772H}, the South Pole Telescope \citep[SPT;][]{2014SPIE.9153E..1PB}, and will continue in the future with the Simons Observatory \citep[SO;][]{2019JCAP...02..056A}, and CMB-S4 \citep{2016arXiv161002743A, 2019arXiv190704473A, 2017arXiv170602464A}.

A key observational diagnostic is the scaling relation between tSZ flux and halo mass. Using a matched multi-filter \citep{2006A&A...459..341M}, \citetalias{2013A&A...557A..52P} reported a nearly self-similar scaling relation for $\sim$260,000 locally brightest galaxies (LBGs) extracted from the Sloan Digital Sky Survey \citep[SDSS;][]{2009ApJS..182..543A}. To convert observable stellar mass to halo mass, they generated a mock catalog using the Millennium simulation, selected LBG samples following the SDSS selection criteria, and derived the stellar mass-halo mass relation. The tSZ signal was measured down to a stellar mass of $\sim 2 \times 10^{11}~\textrm{M}_\odot$, spanning a mass range of almost two orders of magnitude ($M_{500}$ between $\sim 10^{13}$ and $\sim 10^{15}$~$\textrm{M}_\odot$). \cite{2015ApJ...808..151G} later analyzed the data applying aperture photometry instead of the matched filter, taking the stacking bias into account, and obtained results consistent with \citetalias{2013A&A...557A..52P}. \cite{2020ApJ...890..156P} found a self-similar scaling relation for X-ray selected groups and clusters with masses above $10^{13.4} \textrm{M}_\odot$. 

This self-similarity might seem to contradict the intuition that non-gravitational processes, such as stellar and AGN feedback, should cause a mass-dependent deviation from self-similarity with greater deviation at lower mass \cite[e.g.,][]{2015MNRAS.451.3868L, 2016MNRAS.463.4533V}. We will see below that simulations can exhibit self-similarity or deviate from it depending on the radial scale over which quantities are defined. It is worth noting in this context that, because of the large instrumental beam, the \Planck\ result actually measured the tSZ flux on scales out to $5R_{500}$ and extrapolated back to $R_{500}$ using the filter template based on the pressure profile from \cite{2010A&A...517A..92A}.  

In parallel with the observations, the tSZ effect has been studied with several hydrodynamic simulations \citep[e.g.,][]{2015MNRAS.451.3868L, 2017MNRAS.465.2936M, 2017MNRAS.471.1088B, 2018MNRAS.480.4017L, 2021MNRAS.504.5131L, 2019MNRAS.485.3783D}. For example, \cite{2015MNRAS.451.3868L} performed mock tSZ observation using the cosmo-OWLS cosmological simulations. They studied several variations from the `reference' model that included metal-dependent radiative cooling and stellar feedback. Their `non-radiative' model only invoked the UV and X-ray background, without cooling and feedback prescriptions, while the AGN models included AGN feedback with varying temperature treatment. They found that both the tSZ signal and radial pressure profile change substantially among the models. An important implication is that extrapolation from $5R_{500}$ to $R_{500}$ based on a fixed $\beta$ pressure profile, as in the \Planck\ analysis, can significantly bias the results.  Higher angular resolution (arcmin or better), or full forward modeling from simulation to observation, is therefore critical for probing the CGM in galaxy-scale halos. 

\cite{2021MNRAS.504.5131L} compared the tSZ measurements from the \Planck\ data to a number of hydrodynamical simulations. They analyzed the samples within the mass range of $M_{500} \sim 10^{12-14.5}~\textrm{M}_\odot$ and showed that the integrated tSZ flux $\tilde{Y}_{R_{500}}$ of the models (Illustris, TNG-300, EAGLE, and Magneticum) starts to deviate from the self-similar relation below $M_{500} \sim 10^{13} \textrm{M}_\odot$. Their comparison showed that the simulated tSZ flux deviates to a greater extent as halo mass $M_{500}$ decreases, and each simulation gives a different amount of discrepancy from the self-similar relation due to the feedback model they adopted. These preliminary studies demonstrate the potential of SZ observations to add a new kind of constraint on galaxy formation models.  

In this section, we analyze the tSZ signal of the halos in TNG, EAGLE, and FIRE-2 simulations. Since the large-volume cosmological simulations provide sufficient samples for a statistical study, we first explore the mass scaling relation with TNG and EAGLE, especially by separating galaxy types. Then, we compare the radial profiles of the thermodynamic properties for $\sim 10^{12}~\textrm{M}_\odot$ galaxies from the three simulations, including the zoom-in FIRE-2 simulation. Equation~\ref{eq:y_r500} indicates that the tSZ flux $\tilde{Y}_{R_{500}}$ is proportional to $(\Omega_\textrm{b}/\Omega_\textrm{m}) h^{2/3}$ once the self-similar pressure profile is assumed \citep{2010A&A...517A..92A, 2021MNRAS.504.5131L}. Since the \citetalias{2013A&A...557A..52P} analysis and the simulations we use in this analysis adopted different cosmological parameters \citep{2011ApJS..192...18K, 2014A&A...571A..16P, 2016A&A...594A..13P}, we apply the correction to scale the measurements to \citetalias{2013A&A...557A..52P} result.

\subsection{Sample selection in TNG and EAGLE}\label{subsec:sample}

We use simulated galaxies to compare the mass scaling relation with tSZ flux, depending on galaxy type. The previous observational and simulation studies of tSZ flux and halo mass imply the existence of a reliable self-similar relation down to $\sim 10^{13} \textrm{M}_\odot$. To see whether the self-similarity extends to lower masses, we limit our galaxy samples to the halo mass range of $10^{11.0}\textrm{M}_\odot < M_{500} < 10^{13.5}\textrm{M}_\odot$ in order to explore the effect of feedback in the simulations. (Note that the simulation study of \citealt{2021MNRAS.504.5131L} used the mass range of $M_{500} \sim 10^{12-14.5}~\textrm{M}_\odot$.) To understand how galaxy type affects the gas distribution around halos, and thus the tSZ signal, we categorize our galaxy samples from the TNG and EAGLE simulations according to their SFR. In general, we divide galaxies into star-forming and passive (quenched or quiescent galaxies with little SFR in the local universe) systems. Star-forming galaxies tend to reside in halos less massive than passive galaxies at given stellar mass \citep{2016MNRAS.457.3200M}. Our analysis uncovers the dependence of the scaling relation on galaxy type, as well as the relations expected in the low mass regime, where future tSZ observations will explore. Furthermore, the SFR is an indicator of stellar feedback with an assumption that the energy from galactic winds is proportional to the instantaneous SFR \citep[e.g.,][]{2018MNRAS.473.4077P, 2018MNRAS.479.4056W}. Our analysis tests the expectation that the SFR affects the tSZ signal close to the host galaxy.

The relation between stellar mass and SFR has been studied for both TNG and EAGLE \citep{2015MNRAS.450.4486F, 2019MNRAS.485.4817D}. Star-forming galaxies follow a relation in the stellar-mass ($M_\star$)$-$SFR plane \citep[e.g.,][]{2011ApJ...730...61K} known as the star-forming main sequence (SFMS). \citet{2019MNRAS.485.4817D} identified the SFMS in the $M_\star-$SFR plane for the TNG simulations. They fit the SFR of the star-forming samples as a function of the stellar mass,
\begin{equation}
    \log\left(\frac{\langle \textrm{SFR} \rangle_\textrm{sf-ing galaxies}}{\textrm{M}_\odot \textrm{yr}^{-1}}\right) = \alpha (z) \log \left( \frac{M_\star}{\textrm{M}_\odot} \right) + \beta (z),
\end{equation}
where they defined the SFR of a halo as an integration of individual SFR within twice the spherical stellar half-mass radius, and the stellar mass as a sum of all the stellar particles within the same radius. To make a coherent comparison, we use the SFR and the $M_\star$ measured within a 30 (physical) kpc aperture for both TNG and EAGLE, following the common EAGLE analysis \citep[e.g.,][]{2020MNRAS.491.4462D}. 

\begin{figure}[b!]
\plottwo{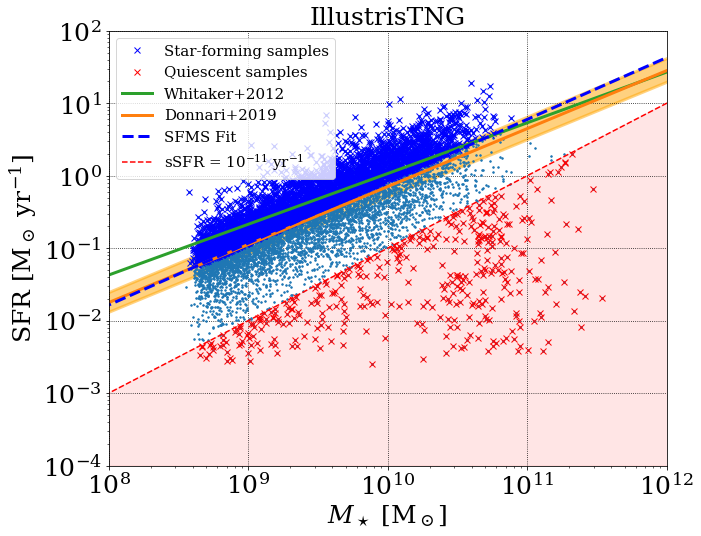}{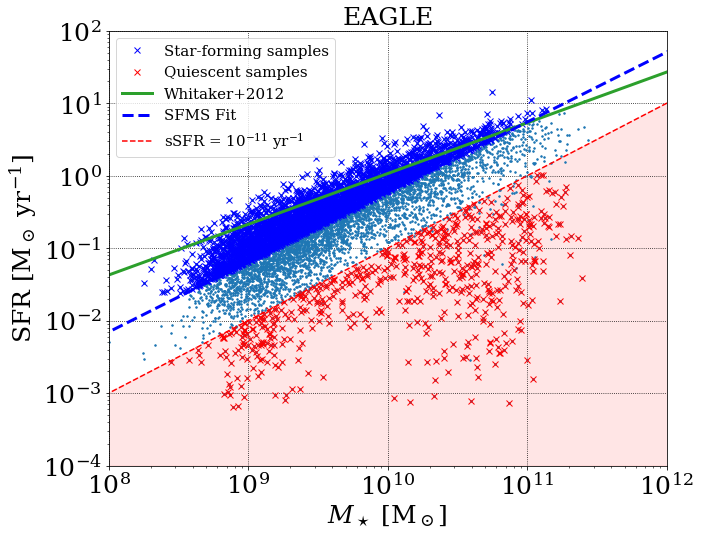}
\caption{Star-forming and quiescent galaxy samples used in our analysis for TNG100-2 (left) and EAGLE Ref-L100N1504 (right) at $z=0$. We use well-resolved central galaxies with $10^{11.0}\textrm{M}_\odot < M_{500} < 10^{13.5}\textrm{M}_\odot$. We designate galaxies above the SFMS fit as star-forming (blue), and those with sSFR $< 10^{-11} \textrm{yr}^{-1}$ as quiescent (red). The galaxies below the SFMS fit but with sSFR $> 10^{-11} \textrm{yr}^{-1}$ are not taken into account to better contrast the effect of galaxy type. The green line shows a fit to the observations \citep{2012ApJ...754L..29W}. The orange line is the SFMS fit to the TNG300, presented in \cite{2019MNRAS.485.4817D} with $\alpha(0) = 0.80 \pm 0.01$ and $\beta(0) = -8.15 \pm 0.11$.}
\label{fig:tsz_sfr_samples}
\end{figure}

We divide galaxies from TNG and EAGLE into star-forming and passive samples. First, we identify the SFMS for central galaxies, whereas \citet{2019MNRAS.485.4817D} did not distinguish centrals from satellites. We consider only systems with more than 50 gas particles and stellar particles each to ensure that the galaxies are well-resolved. Then, we classify as quiescent those galaxies with sSFR below $10^{-11} \textrm{yr}^{-1}$ \citep[e.g.,][]{2013MNRAS.432..336W}, where $\textrm{sSFR} = \textrm{SFR} / M_\star$. We fit the SFMS using the rest of the galaxies, performing a linear fit to the median SFRs of $M_\star$ in the range between $10^{9}$ and $10^{10}$, separated by 0.2 dex bins. We define the galaxies above the MS fit as star-forming samples in this analysis to better contrast the results depending on galaxy types.

Figure~\ref{fig:tsz_sfr_samples} shows the $M_\star-$SFR plane populated with the present-day central galaxies in TNG (snapshot 99) and EAGLE (snapshot 28), and our aforementioned samples. The criteria we use give 4,178 star-forming and 1,590 quiescent galaxies for TNG, and 3,355 star-forming and 907 quiescent galaxies for EAGLE. Since TNG has $\sim$30\% larger volume than EAGLE, it accordingly contains more galaxies. However, these are of similar order and will not affect the interpretation.

\subsection{Integrated tSZ signal: Mass scaling relation}\label{subsec:tsz_integrated}

In this section, we examine the tSZ mass scaling relation ($Y\!-\!M$ relation) for the TNG and EAGLE galaxy samples, divided by SFR. The aims of showing the $Y\!-\!M$ relation using the simulated galaxy samples are to: (1) compare results to the \citetalias{2013A&A...557A..52P} data, where the relation shows little departure from a single power-law with a slope of 5/3 and, (2) investigate how the relation depends on the type of galaxy (star-forming vs. quiescent) and understand how the feedback physics included in each simulation can affect the relation.

\begin{figure}[b!]
\plottwo{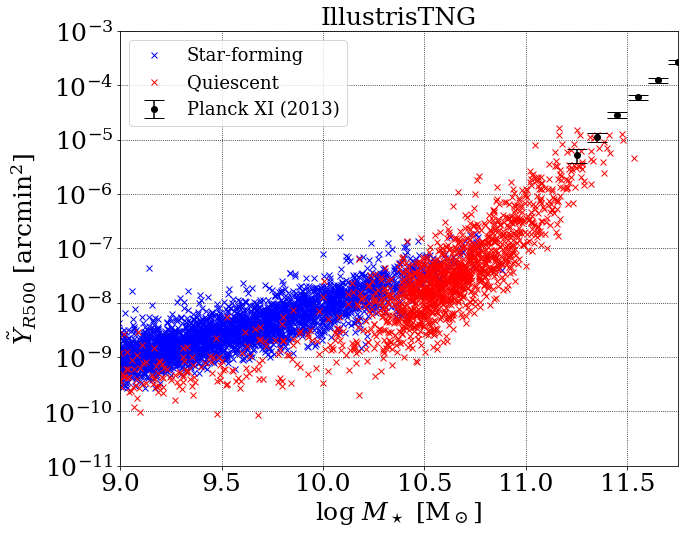}{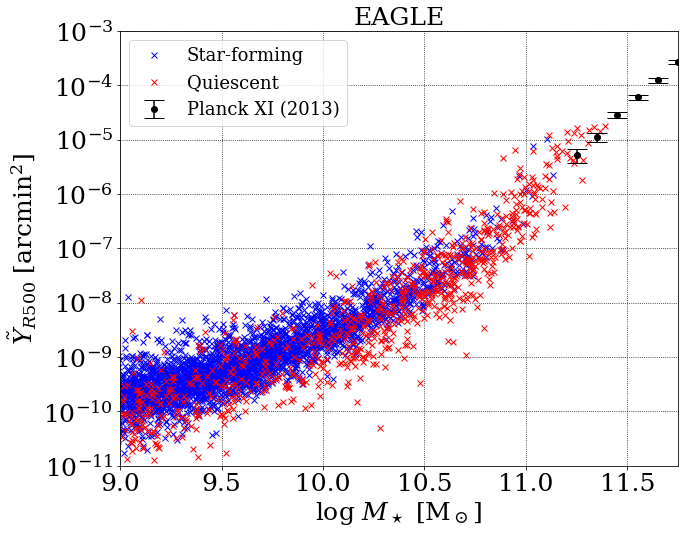}
\caption{tSZ flux as a function of stellar mass for TNG (left) and EAGLE (right). Blue and red crosses indicate the star-forming and quiescent samples as defined in Figure~\ref{fig:tsz_sfr_samples}. Black circles show the \citetalias{2013A&A...557A..52P} measurements, and the error bars indicate the statistical error. Note that our samples do not cover the entire range of the stellar mass from the \citetalias{2013A&A...557A..52P} observation since we limited our galaxy mass range to $10^{11.0}\textrm{M}_\odot < M_{500} < 10^{13.5}\textrm{M}_\odot$.}
\label{fig:tsz_mstar}
\end{figure}

\begin{figure}[t!]
\plottwo{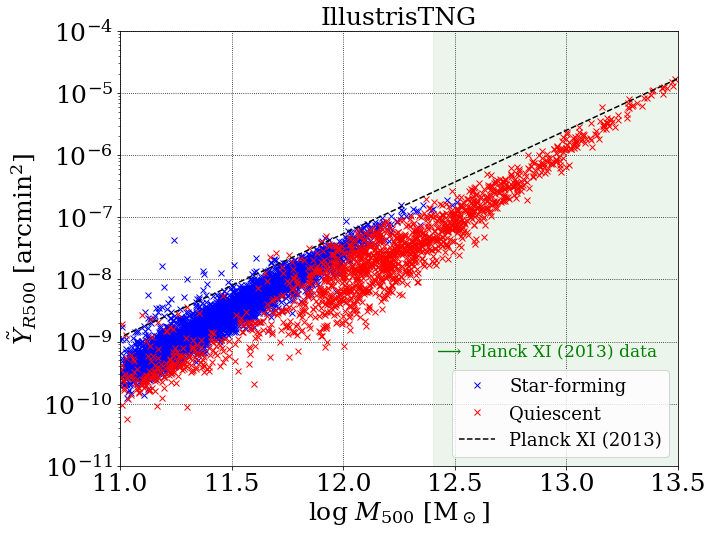}{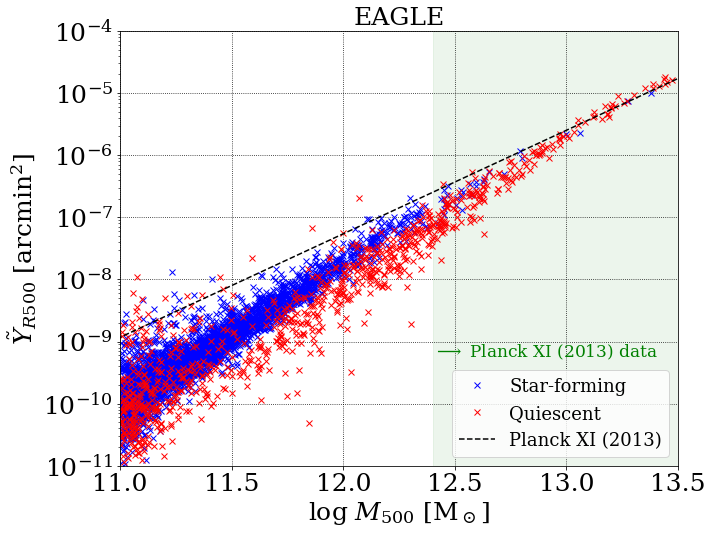}\\\vspace{0.2in}
\plottwo{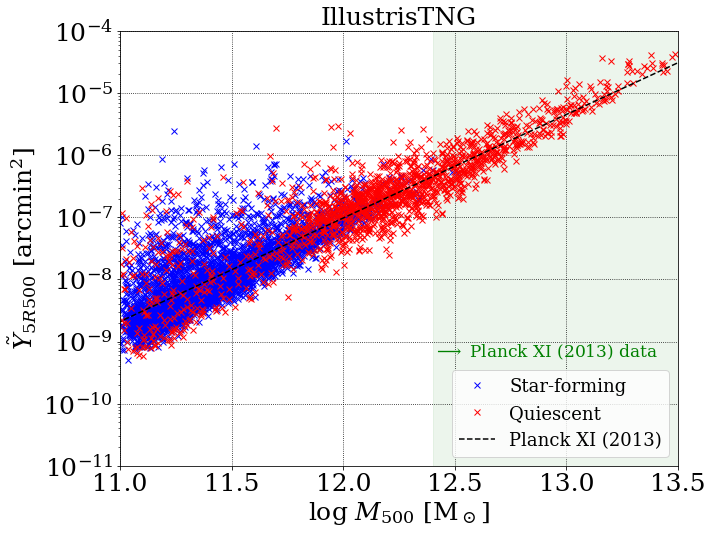}{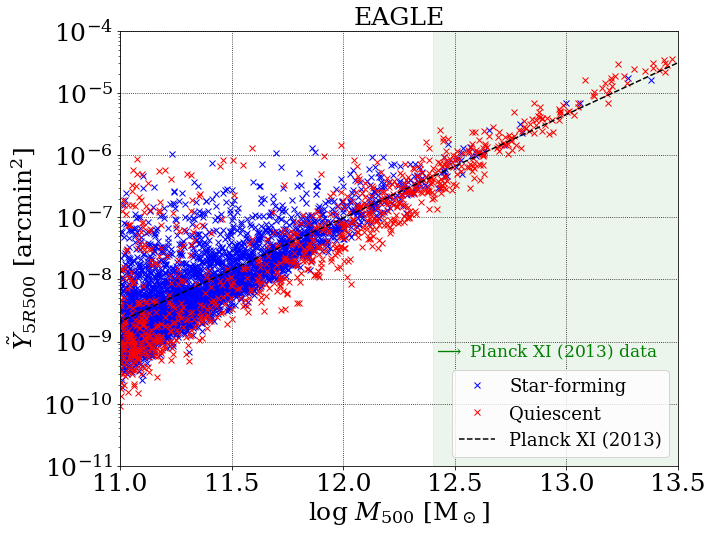}
\caption{tSZ flux as a function of halo mass $M_{500}$, integrated out to $R_{500}$ (upper), and to $5 R_{500}$ (lower) in TNG (left column) and EAGLE (right column). The shaded green region shows the mass range of the galaxy samples used in the \citetalias{2013A&A...557A..52P} analysis, and their best-fit scaling relation is shown as the dotted line. The relation in the lower panels is rescaled by multiplying the original relation in the upper panels by 1.796 (ratio from the universal pressure profile).}
\label{fig:tsz_m500}
\end{figure}

We spherically integrate the tSZ signal out to $R_{500}$ to calculate $Y_{R_{500}}$ for each of the galaxy samples. The $Y\!-\!M$ relation is often presented as a function of the halo mass $M_{500}$ to relate the observable tSZ flux to the dark matter halo properties. However, stellar mass is the more directly measurable property, and observations determine the tSZ to stellar mass relation. The stellar mass$-$halo mass conversion is highly dependent on galaxy properties, especially on color. Figure~\ref{fig:tsz_mstar} shows the spherically integrated $Y_{R_{500}}$ as a function of $M_\star$, with the \citetalias{2013A&A...557A..52P} measurements overlaid. The simulations and the \citetalias{2013A&A...557A..52P} observation overlap around $10^{11.25}\textrm{M}_\odot < M_{*} < 10^{11.5}\textrm{M}_\odot$. Note that the \citetalias{2013A&A...557A..52P} points in their work flatten out below $M_\star \sim 10^{11}~\textrm{M}_\odot$ because of dust contamination. The $M_\star = 10^{11.25}\textrm{M}_\odot$ bin of the \Planck\ measurement was made at 3.5$\sigma$ detection, and the lower mass bins show weaker detection significance, therefore we only include the points above $M_\star = 10^{11.25}\textrm{M}_\odot$ in Figure~\ref{fig:tsz_m500}.

\begin{figure}[t!]
\plottwo{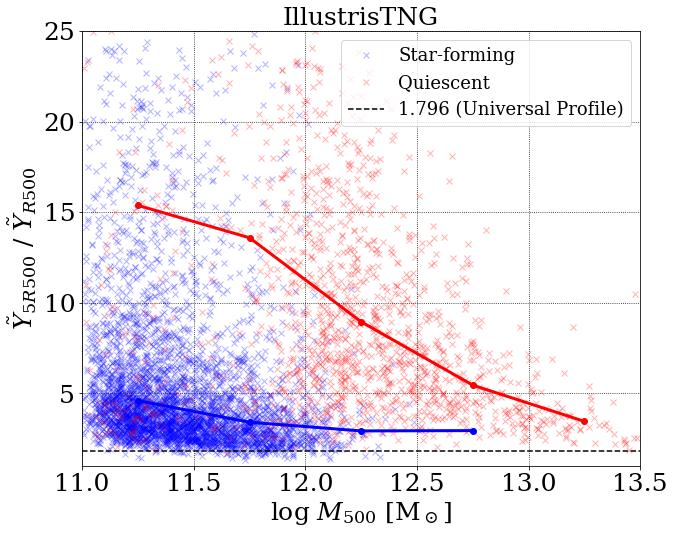}{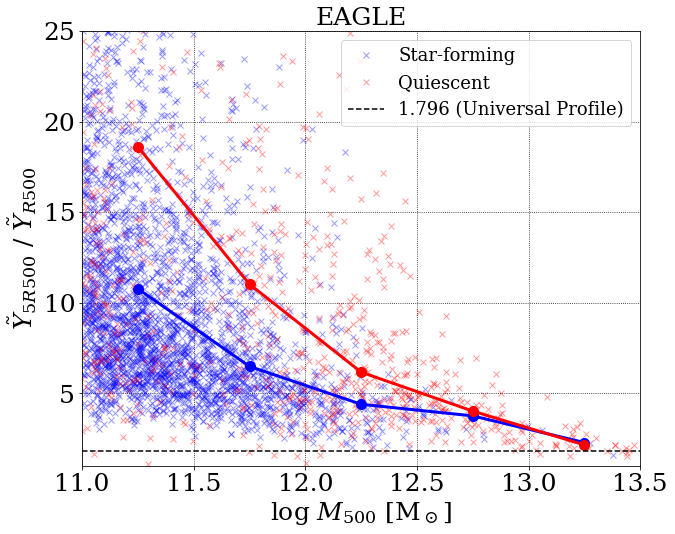}
\caption{$Y_{5R_{500}} / Y_{R_{500}}$ of the TNG and EAGLE galaxies. The dashed line shows the conversion factor 1.796 corresponding to the universal pressure profile. The red and blue dots connected by the solid lines indicate the median value of the ratios in each $M_{500}$ mass bin. The mass bins are separated by 0.5 dex between $10^{10.5} \textrm{M}_\odot$ and $10^{13.5} \textrm{M}_\odot$. Higher ratios of $Y_{5R_{500}} / Y_{R_{500}}$ indicate that the hot gas has been pushed out to the outskirts of the galaxy halos by feedback.}
\label{fig:tsz_yratio}
\end{figure}

Figure~\ref{fig:tsz_m500} shows the simulated tSZ flux as a function of halo mass $M_{500}$ and compares  to the \citetalias{2013A&A...557A..52P} scaling relation. As described earlier, \citetalias{2013A&A...557A..52P} measured the tSZ signal out to $5 R_{500}$ and converted it to $Y_{R_{500}}$ assuming the universal pressure profile (UPP; Equation~22 in \citealt{2010A&A...517A..92A}). This rescaling of the tSZ flux to go from $Y_{R_{500}}$ to $Y_{5R_{500}}$ corresponds to a factor of 1.796. We present the scaling relation in terms of both $Y_{R_{500}}$ and $Y_{5R_{500}}$. The simulated galaxies lie below the \citetalias{2013A&A...557A..52P} scaling relation for $Y_{R_{500}}\!-\!M_{500}$, and that the discrepancy grows as the mass decreases for both types of galaxies and for both simulations. If, on the other hand, we consider the scaling relation in $Y_{5R_{500}}\!-\!M_{500}$, we find good agreement.

To test the explanation of this `aperture size' effect in the tSZ flux measurement, we show the ratio $Y_{5R_{500}} / Y_{R_{500}}$ as a function of halo mass and separated by galaxy type in Figure~\ref{fig:tsz_yratio}. The stark difference between these two scaling relations ($Y_{R_{500}}\!-\!M_{500}$ and $Y_{5R_{500}}\!-\!M_{500}$) could be explained qualitatively by feedback driving gas from small radii ($\sim R_{500}$) out to larger radii such that all the gas and energy one expects are contained inside $5R_{500}$ but not inside $R_{500}$. We observe two important trends. The first is that, for all galaxies, the ratio converges to the universal pressure profile value as the halo mass increases. The second is that the discrepancy with the UPP is larger for quiescent galaxies. Both of these trends support the feedback explanation. The trend with mass suggests that, as the halo mass grows, gravity wins out over feedback effects driving gas outward, resulting in better agreement with the UPP (derived from massive clusters, see below), or, equivalently, feedback is increasingly effective in overcoming gravity at $R_{500}$ as the halo mass decreases. The observation of a larger discrepancy for quiescent galaxies is consistent with the idea that feedback has had its full effect in that class, blowing the bulk of the gas fuel for star formation out into the far CGM, while star-forming galaxies are still somewhere along that evolutionary trend, with less of the gas blown out past $R_{500}$, making it possible for them to continue forming stars at a high rate.

These deviations from self-similarity at $R_{500}$ occur simply because the UPP, on which the extrapolation from $Y_{5R_{500}}$ to $Y_{R_{500}}$ is based, is not correct for galaxies.  One must remember that it was obtained from a sample of massive clusters, in which feedback plays a much smaller role (fractionally). In the mass range for which they overlap, $10^{11.5} < M_{500} < 10^{12.5}$, the relative locations of star-forming and quiescent galaxies in Figures~\ref{fig:tsz_m500} and \ref{fig:tsz_yratio} is also illuminating. The much larger $Y_{5R_{500}} / Y_{R_{500}}$ ratio for quiescent galaxies suggests that a much larger fraction of their gas has been heated and pushed out to large radius than for star-forming galaxies. This is consistent with the idea that star formation in quiescent galaxies was quenched by feedback from star formation and AGN activity that caused this heating/pushing of the gas. (Though the details of the model, including the timescale it takes to set up this inequality in the profile, need to be further studied, and it is outside the scope of this work.) Additionally, the growth of the ratio and the difference between star-forming and quiescent galaxies at smaller masses indicates that the effect is fractionally larger at lower masses.  While the details are quite dependent on the simulation, the effect is clearly present in both simulations and is consistent with expectations from the basic picture of feedback quenching star-forming galaxies by driving their gas out to large radius and heating it.

\subsection{Comparison with the ACT stacked profile}

\begin{figure}[t!]
\plottwo{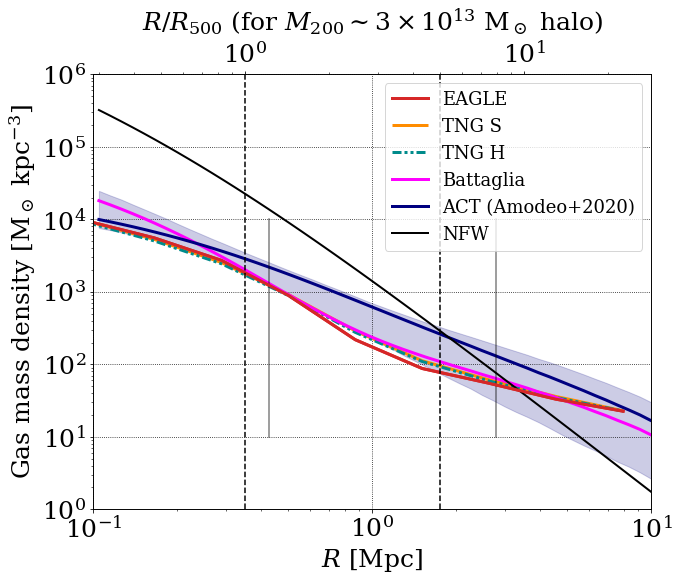}{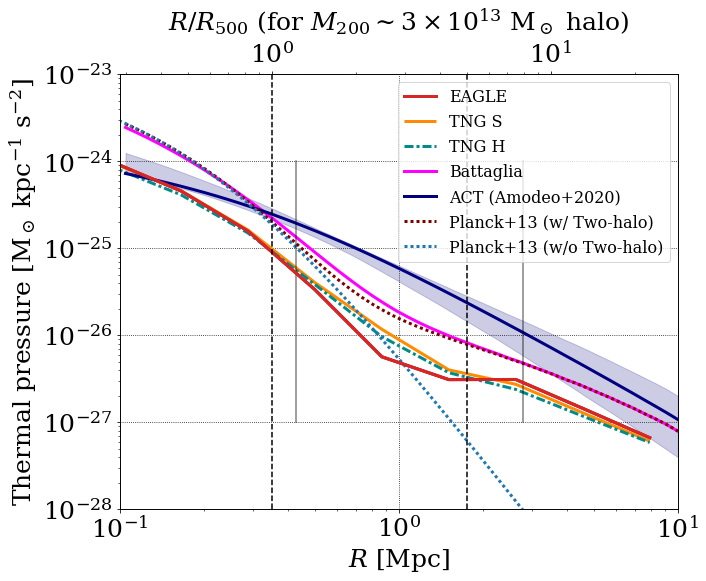}
\caption{Deprojected radial gas mass density (left) and thermal pressure (right) profiles from the ACT and CMASS cross correlation \citep[][blue line and band]{2021PhRvD.103f3514A} compared TNG (orange and green), EAGLE (red), \cite{2010ApJ...725...91B} simulation (magenta), and the NFW profile \citep[][black]{1997ApJ...490..493N}. The dotted line in the thermal pressure plot shows the best-fit GNFW pressure profile derived from the galaxy cluster observations in \cite{2013A&A...550A.131P}. We show both the original pressure profile without a two-halo term (dotted blue), and with a two halo term (dotted maroon) presented by \cite{2021PhRvD.103f3514A}. The upper axis shows the radial distance in units of $R_{500} = 350$~kpc, for $M_\textrm{200} \sim 3 \times 10^{13} \textrm{M}_\odot$ halo. The dashed vertical lines show $R=R_{500}$ and $R=5R_{500}$. The vertical grey bars indicate the radial ranges where the ACT SZ measurements were made \citep{2021PhRvD.103f3513S}.}
\label{fig:eagle_act2020tsz}
\end{figure}

Recently, the ACT collaboration presented stacked tSZ and kSZ measurements by cross-correlating the ACT DR5 dataset and the BOSS CMASS galaxy catalog \citep{2021PhRvD.103f3513S, 2021PhRvD.103f3514A}. From the SZ measurements, they inferred the gas mass density and temperature, hence thermal pressure, profiles by fitting a parameterized general Navarro-Frenk-White (NFW) profile \citep{2012ApJ...758...74B}. They compared the results to the NFW profile \citep{1997ApJ...490..493N}, hydrodynamical simulations by \cite{2010ApJ...725...91B}, and TNG galaxy samples as shown in Figure~\ref{fig:eagle_act2020tsz} (Figure~6 in \citealt{2021PhRvD.103f3514A}). Although the simulations matched the profiles deprojected from the observations well at radii smaller than $\sim 2 R_{200}$ ($\sim$ 1~Mpc), the discrepancy was not insignificant at larger radii, especially for the thermal pressure profile, thus temperature. They claimed the simulations under-predicted the gas density and thermal pressure, and explained that it was due to the CGM gas in the outer region being less heated by the sub-grid models the simulations employed. In this section, we present the stacked radial profiles using the EAGLE galaxies and compare those to the ACT best-fit results and the TNG profiles in \cite{2021PhRvD.103f3514A}.

The CMASS galaxies are distributed within the redshift range $0.4 < z < 0.7$, and the galaxy samples used in the ACT analysis have median redshift $z=0.55$ and mean stellar mass of $3 \times 10^{11} \textrm{M}_\odot$ (from \citealt{2013MNRAS.435.2764M} stellar mass estimates) that corresponds to a halo mass $M_\textrm{200} \sim 3 \times 10^{13} \textrm{M}_\odot$ using the \cite{2018AstL...44....8K} stellar-to-halo mass conversion. Note that \cite{2021PhRvD.103f3514A} used $R_{200}$ and $M_{200}$ as virial radius and mass, whereas we use $R_{500}$ and $M_{500}$ throughout this paper. For their $M_\textrm{200} \sim 3 \times 10^{13} \textrm{M}_\odot$ halo, $R_{200}$ and $R_{500}$ correspond to $\sim 500$~kpc and $\sim 350$~kpc, respectively. They selected red galaxies in the TNG simulation and calculated the average profile by weighting the samples using the stellar and halo mass distributions of the observed galaxies (`TNG S' and `TNG H' in Figure~\ref{fig:eagle_act2020tsz}). For example, they used nine log-spaced bins between $10^{11.53}$ and $10^{13.98}~\textrm{M}_\odot$ for the halo mass distribution. The mass catalog can be downloaded as part of the \texttt{Mop-c GT} (``Model-to-observable projection code for Galaxy Thermodynamics'') package. We use the identical color cut ($g - r \geq 0.6$)\footnote{$g$ and $r$ are the Sloan Digital Sky Survey (SDSS) magnitudes.} for the EAGLE central galaxies, using snapshots 22 ($z=0.62$) and 23 ($z=0.50$) to match the median redshift. We calculate the weighted average of the gas mass and thermal pressure profiles using the halo mass ($M_{200}$) probability density function of \cite{2021PhRvD.103f3514A}. Since the gas mass density and the pressure are proportional to $M$ and $M^{5/3}$, respectively, it is necessary to use the correct mass distribution to estimate the stacked profiles \citep{2021ApJ...919....2M}. As described earlier, we include the two-halo term in calculating the radial profiles by taking the nearby particles within a volume into account.

Figure~\ref{fig:eagle_act2020tsz} shows the gas mass density and thermal pressure profiles inferred from the EAGLE data, compared to the ACT best-fit profiles as well as the TNG results in \cite{2021PhRvD.103f3514A}. The EAGLE density and pressure profiles are nearly consistent with the TNG profiles. Both the EAGLE and TNG density profiles are within 2$\sigma$ of the ACT profile at most radial distances. The thermal pressure profiles show somewhat similar discrepancies among ACT, TNG, and EAGLE. 

We could interpret the comparison of the simulation data (EAGLE, TNG) to the ACT profiles in two ways. First, the EAGLE profiles are approximately consistent with the TNG profiles. This supports the discussion in \cite{2021PhRvD.103f3514A} that the sub-grid models in the cosmological simulations under-predict the gas density and pressure at large radii due to insufficient heating. However, as we will see in Section~\ref{subsec:radial}, the FIRE-2 simulation with the cosmic-ray (CR) treatment predicts even lower thermal pressure than TNG and EAGLE in general, primarily due to the temperature than the gas mass density, and the most recent FIRE-2 sub-grid model will not likely resolve this issue. Another possible interpretation is that feedback in TNG and EAGLE is too effective, that they blow too much gas out to a large radius. We argued earlier that feedback in TNG and EAGLE was not as effective in star-forming galaxies as in quiescent galaxies (Section~\ref{subsec:tsz_integrated}). As we will see in Figure~\ref{fig:mw_radial}, the density and pressure for star-forming galaxies are higher than for quiescent galaxies in both simulations. This implies that perhaps the low gas mass density and pressure profiles of TNG and EAGLE are results of very efficient feedback mechanisms.

Also, we could relate the comparison to the integrated tSZ flux and mass scaling relation in Section~\ref{subsec:tsz_integrated}. Figure~\ref{fig:tsz_m500} shows that the simulated tSZ flux integrated out to $5R_{500}$ more or less agree with the \citetalias{2013A&A...557A..52P} fit around the mean halo mass of the galaxy samples ACT analysis used ($M_\textrm{200} \sim 3 \times 10^{13} \textrm{M}_\odot$ or $M_\textrm{500} \sim 2 \times 10^{13} \textrm{M}_\odot$). However, if we use the ACT pressure profile in Figure~\ref{fig:eagle_act2020tsz} to calculate the tSZ flux out to $5R_{500}$, it is a few times higher than the \citetalias{2013A&A...557A..52P} tSZ flux measurement. The discrepancy is dependent on the two-halo term and could be due to the sample selection.  For example, \citetalias{2013A&A...557A..52P} explicitly selected `locally brightest galaxies' to construct a sample of central galaxies, which is defined as the galaxies that don't have a brighter neighbor within 1~Mpc. An investigation into this effect would require fully modeling the selection functions of the \citetalias{2013A&A...557A..52P} and CMASS samples, which is beyond the scope of this work.

\subsection{Gas radial profiles of Milky Way sized galaxies}\label{subsec:radial}

The study of integral quantities in Section~\ref{subsec:tsz_integrated} strongly suggests that there is a great deal of information about the impact of feedback on the CGM in the tSZ radial profile. Future SZ observations will be able to characterize the tSZ signal from Milky Way-sized or even less massive halos through stacking if individual detections are not possible. The CGM properties of Milky Way-mass halos have been studied with cosmological simulations. For example, \cite{2016MNRAS.462.3751K, 2019MNRAS.486.4686K} selected Milky Way-mass galaxies at $z=0$ in Illustris and TNG simulations, based on both halo and stellar masses and explored the CGM structure, including the radial distribution of the gas. In this section, we explore the radial profiles of TNG and EAGLE galaxies in the mass range $10^{11.75} < M_{500} < 10^{12.25} \textrm{M}_\odot$, separated by galaxy type, and we also include FIRE-2 simulations of individual Milky Way-sized galaxies (see Table~\ref{table:fire2_list}).

\begin{deluxetable}{ccccccc}[b!]
\tablecolumns{8}
\tablewidth{0pc}
\tablecaption{FIRE-2 Simulation Summary}
\tablehead{\colhead{Name} & \colhead{$M^\textrm{vir}_\textrm{halo} (\textrm{M}_\odot)$} & \colhead{$M_* (\textrm{M}_\odot)$} & \colhead{$R_\textrm{vir}$ (kpc)} & \colhead{$M_{500} (\textrm{M}_\odot)$} & \colhead{$R_{500}$ (kpc)}}
\startdata
m12f & $1.6 \times 10^{12}$ & $8.0 \times 10^{10}$ & 306 & $1.2 \times 10^{12}$ & 163\\
m12i & $1.2 \times 10^{12}$ & $6.5 \times 10^{10}$ & 275 & $9.1 \times 10^{11}$ & 150\\
m12m & $1.5 \times 10^{12}$ & $1.2 \times 10^{11}$ & 301 & $1.1 \times 10^{12}$ & 159\\
\enddata
\tablecomments{List of the FIRE-2 zoom-in simulations used in this work. The physical quantities are measured in the reference simulation dataset \citep[see][for details]{2018MNRAS.480..800H}. We used both the reference and CR runs for each of the models.}
\label{table:fire2_list}
\end{deluxetable}

\begin{figure}[t!]
    \begin{center}
    \includegraphics[width=.33\textwidth]{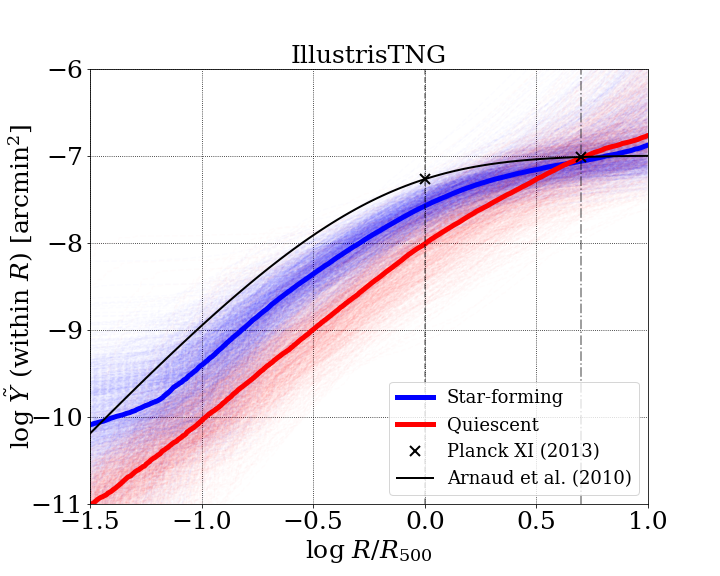}\hfill
    \includegraphics[width=.33\textwidth]{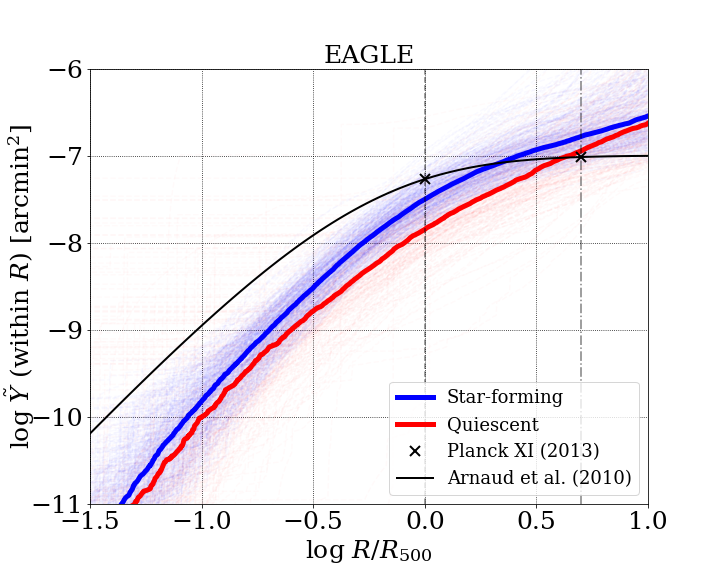}\hfill
    \includegraphics[width=.33\textwidth]{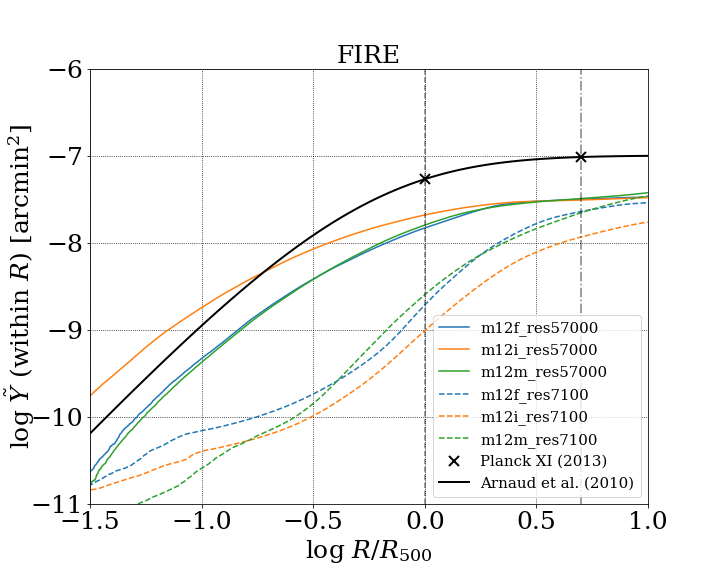}\\
    \includegraphics[width=.33\textwidth]{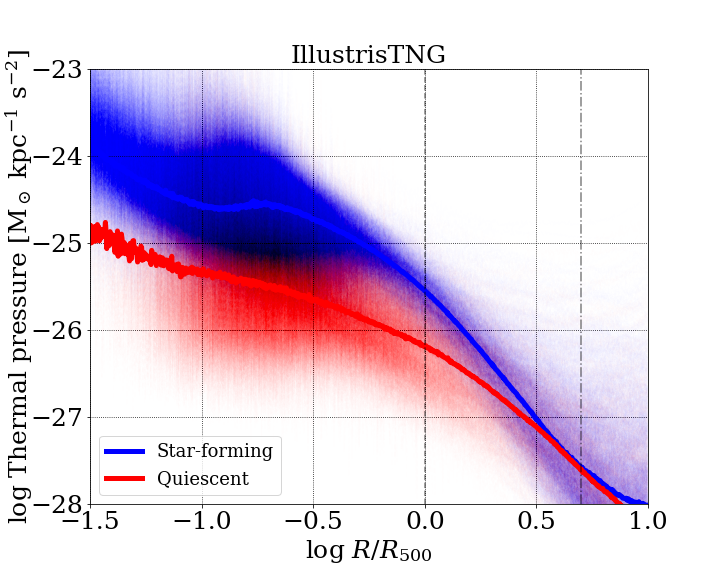}\hfill
    \includegraphics[width=.33\textwidth]{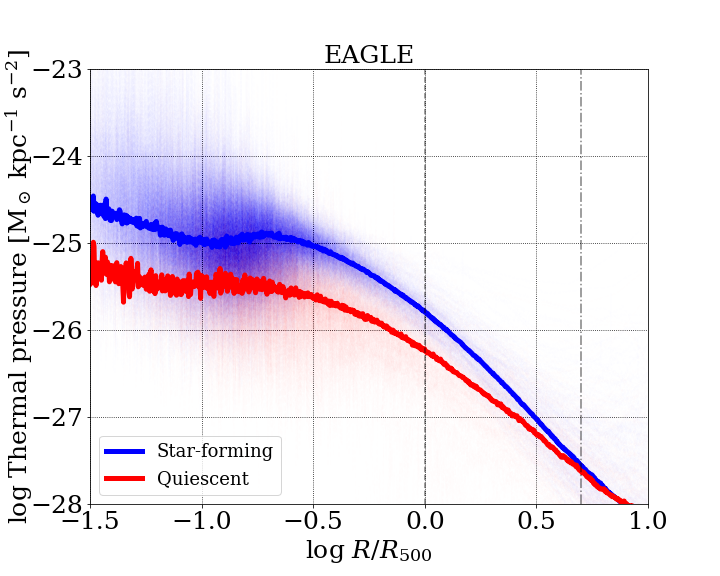}\hfill
    \includegraphics[width=.33\textwidth]{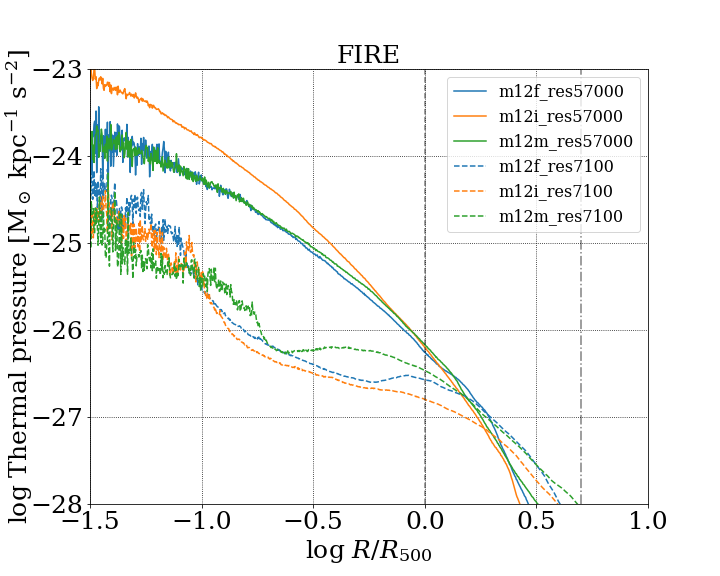}\\
    \includegraphics[width=.33\textwidth]{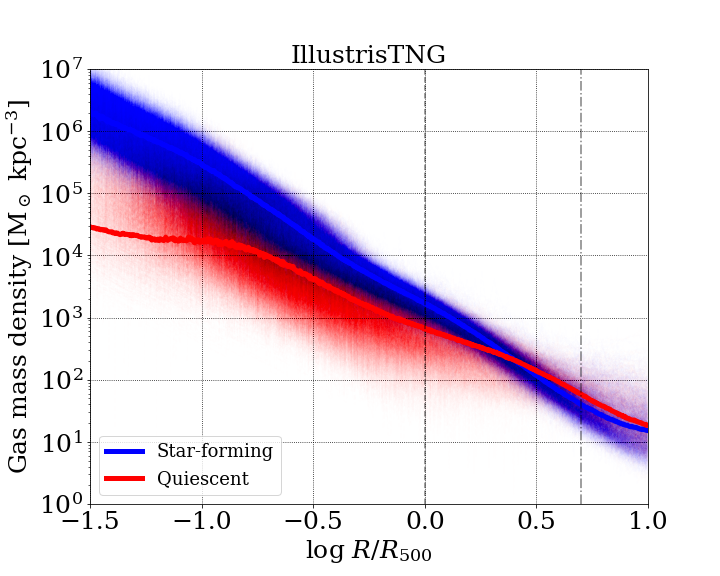}\hfill
    \includegraphics[width=.33\textwidth]{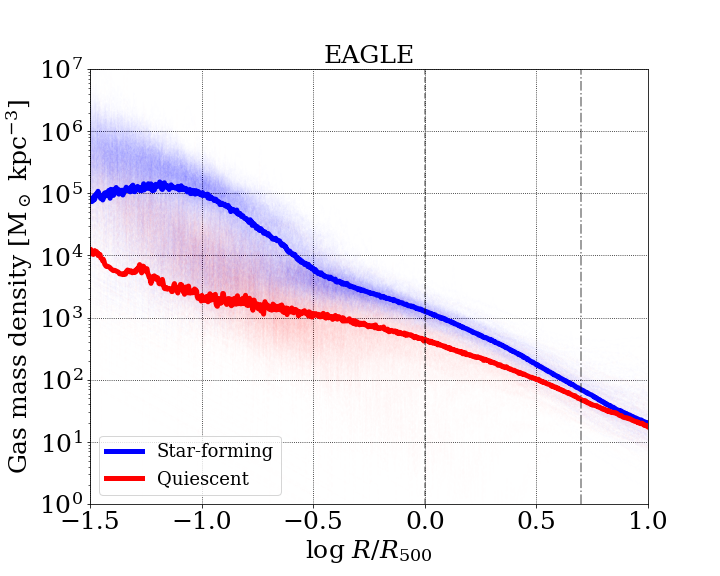}\hfill
    \includegraphics[width=.33\textwidth]{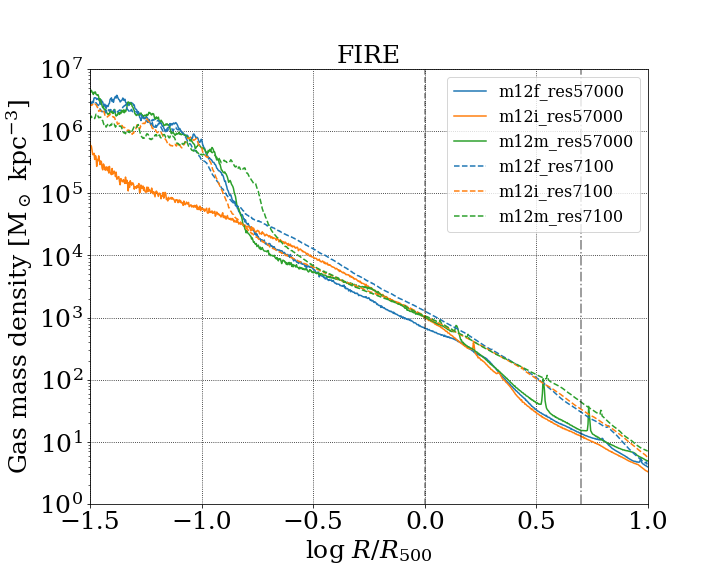}\\
    \includegraphics[width=.33\textwidth]{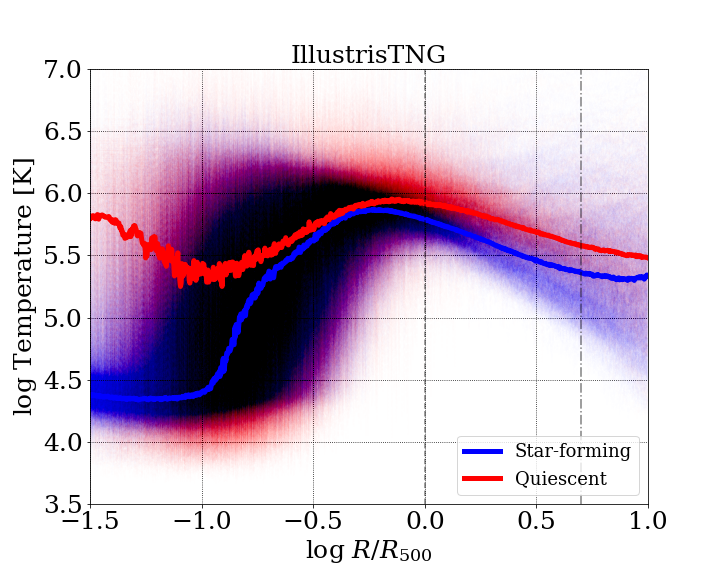}\hfill
    \includegraphics[width=.33\textwidth]{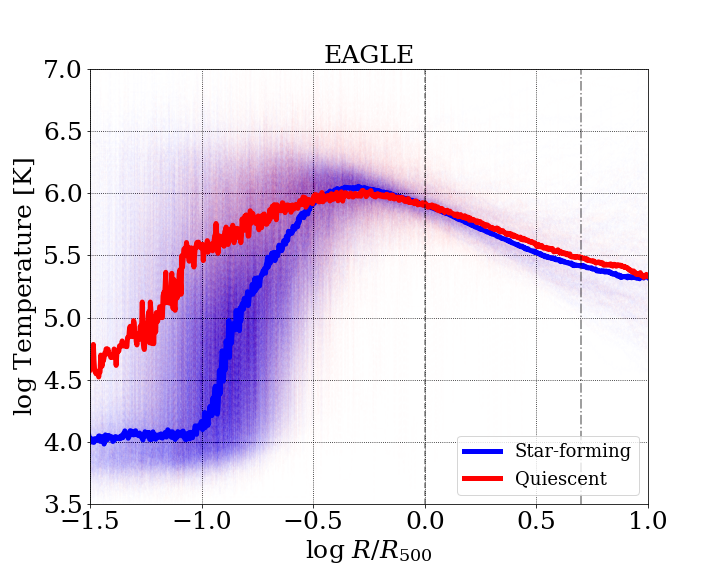}\hfill
    \includegraphics[width=.33\textwidth]{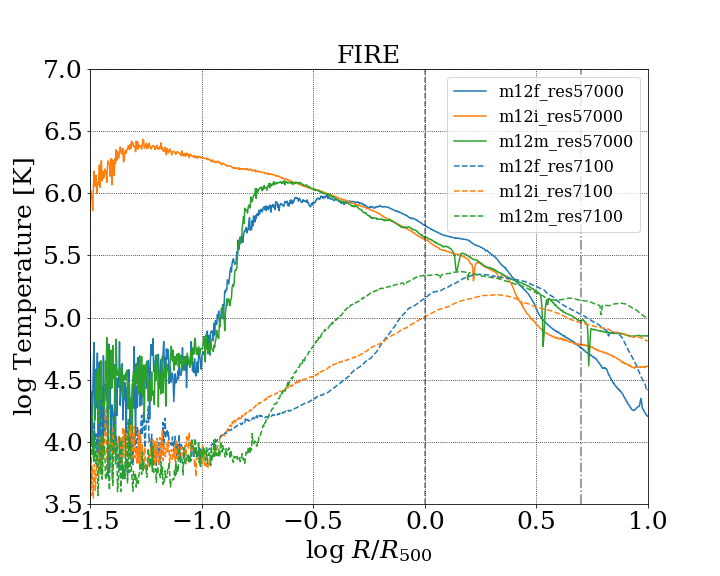}
    \caption{Mean radial profiles of the TNG (left column) and EAGLE (middle column) galaxies separated by SFR in the mass range $10^{11.75} < M_{500} < 10^{12.25} \textrm{M}_\odot$: Blue and red curves show median radial profiles of the star-forming and quiescent samples, respectively. We show cumulative tSZ flux, thermal pressure, gas mass density, and temperature from top to bottom. The faint dotted lines are the radial profiles of the individual galaxies. The black crosses in the tSZ radial profile plots (top row) denote the tSZ signal for $10^{12} \textrm{M}_\odot$ mass halo at $R_{500}$ ($\sim 160$~kpc) and $5 R_{500}$, inferred from the self-similar relation of \citetalias{2013A&A...557A..52P}, whose measurements are based on the halos at least an order of magnitude more massive than the MW-sized halo. The solid black lines in those plots show spherically integrated tSZ flux as a function of radius, derived from the universal pressure profile \citep{2010A&A...517A..92A}, normalized by the \citetalias{2013A&A...557A..52P} measurement. Radial profiles of the FIRE-2 galaxies (right column): The solid and dashed lines show the reference models and the CR included simulations, respectively. The FIRE-2 m12i, m12m, and m12f are all `star-forming galaxies' as opposed to quiescent.}
    \label{fig:mw_radial}
    \end{center}
\end{figure}

Figure~\ref{fig:mw_radial} shows the mass-weighted radial profiles, including cumulative tSZ signal, gas mass density, temperature, and thermal pressure. We selected the star-forming and quiescent samples in TNG and EAGLE based on the sample selection in Section~\ref{subsec:sample}, and limited the halo mass ($M_{500}$) range to $10^{11.75 - 12.25}\textrm{M}_\odot$. This selection leaves 602 star-forming and 699 quiescent galaxies for the TNG and 392 star-forming and 226 quiescent galaxies for the EAGLE dataset. Each of the dashed lines in the background shows an individual galaxy radial profile, and the solid, thick curves display the median profiles. When calculating the radial profiles, we include all the gas particles around the center of each halo. However, a distinction between ISM and CGM gas particles can be made by using the SFR of the fluid elements \citep{2019MNRAS.485.3783D, 2020MNRAS.491.4462D}. Since we mainly focus on the thermodynamic profiles around and beyond the virial radius, the separation of the ISM particles has a negligible effect on our results.

As shown in Section~\ref{subsec:tsz_integrated}, star-forming galaxies tend to exhibit higher tSZ fluxes than quiescent galaxies when integrated out to $R_{500}$, although the difference is smaller or even negligible at $5R_{500}$ in both TNG and EAGLE (Figure~\ref{fig:tsz_m500}). This is consistent with the tSZ flux inferred for the $10^{12}\textrm{M}_\odot$ halo at $R=5 R_{500}$ from the \citetalias{2013A&A...557A..52P} measurement (Figure~\ref{fig:mw_radial}). Again, the radial profiles reflect the intuition that the heating and expulsion of gas far out into the CGM has been much more extensive for quiescent galaxies than for star-forming ones. The gas density is larger, and the temperature is lower close in ($\log (R/R_{500}) < -0.5$) for star-forming galaxies compared to  quiescent galaxies, which supplies a reservoir of fresh fuel for star formation. Of course, because the thermal pressure (thus electron pressure) is proportional to the product of the gas density and temperature, these two effects can partially or completely cancel each other. At radii smaller than $\sim 2R_{500}$ in TNG, and at all radii in EAGLE, the gas density wins out over the temperature, yielding higher pressure for star-forming galaxies. At radii larger than $\sim 2R_{500}$ in TNG, the gas density of the quiescent galaxies surpasses that of star-forming galaxies. This leads to higher thermal pressure for $R \gtrsim 2R_{500}$, and the tSZ fluxes of the quiescent and star-forming samples become comparable at $3R_{500}$.

The zoom-in FIRE-2 simulation \citep{2018MNRAS.480..800H} uses better spatial and mass resolutions than the TNG and EAGLE datasets, focusing on the physical processes of individual halos. \cite{2016MNRAS.463.4533V} used the FIRE-1 simulation, which only included the stellar feedback prescription without AGN feedback to study the tSZ effect and the X-ray emission properties of the halos. Incorporating the galaxy simulations' redshift-dependent snapshots, they discovered that the halo's hot gas fraction is dependent on the halo mass, with little redshift dependence. They pointed out that the tSZ measurements could observe this, with increasing suppression of the tSZ signal compared to the \citetalias{2013A&A...557A..52P} self-similar relation as the halo mass decreases below $10^{13} \textrm{M}_\odot$, as a result of stellar feedback. Recently, the FIRE project presented the FIRE-2 simulations with a CR treatment \citep{2020MNRAS.492.3465H, 2020MNRAS.496.4221J}. \cite{2020MNRAS.496.4221J} showed that the non-thermal CR pressure could dominate over thermal pressure in the CGM, supporting the gas, especially for Milky Way-sized halos. In these simulations, gas cooling is more effective than CR heating, and the gas temperatures are expected to decrease considerably compared to the reference simulations, while the gas mass density profiles are not strongly affected. Since the tSZ directly measures the integrated thermal pressure, comparing the FIRE-2 reference and CR runs illustrates how the tSZ flux changes with the different sub-grid models.

The radial profiles of the $z=0$ (snapshot number 600) reference and CR simulation models are shown in the right column of Figure~\ref{fig:mw_radial}. We first observe that the reference simulations and CR runs show substantially different radial profiles, except for the gas mass density profile, as already discussed in \cite{2020MNRAS.496.4221J}. This implies that these CR-driven differences should be considered meaningful. CRs cause a substantial reduction in pressure and temperature out to $3R_{500}$ because they transport energy and momentum created by star formation out to $3R_{500}$ before slowing down and depositing their energy. We also see that the CRs greatly suppress the thermal pressure due to the decreased temperature, resulting in much lower tSZ flux than the reference runs at $R_{500}$. 

When we compare the FIRE-2 to TNG and EAGLE, the FIRE-2 reference simulation's radial profiles in all four quantities ($Y$, gas mass density, pressure, and temperature) are generally in qualitative agreement with TNG and EAGLE. However, we cannot directly compare the FIRE-2 radial profiles to TNG and EAGLE at radii beyond $\sim 1.5R_{500}$ because the zoom-in FIRE-2 simulations do not include the two-halo term from the gas particles outside their simulation volume. We observe that the tSZ radial profiles of the reference simulations flatten for $R \gtrsim 2R_{500}$. These tSZ profiles are similar to the TNG and EAGLE tSZ profiles when the two-halo term is not considered (Figure~\ref{fig:appendix_twohalo}).

\subsection{Effect of AGN feedback in the EAGLE simulation}

We compare the EAGLE simulations Ref-L0050N0752 (`Reference') and NoAGN-L0050N0752 (`NoAGN' simulation) at $z=0$ to explore the effect of the AGN feedback. They are in smaller $50^3$~cMpc$^3$ volumes and use the same particle resolution as Ref-L100N1504. NoAGN simulation does not include the sub-grid model describing the AGN feedback. \cite{2019MNRAS.485.3783D} already compared the simulations by looking at the present-day halo baryon fraction, which is a sum of the gas and stellar mass over the total mass within the radius $R_{200}$. As expected, the halos more massive than $\sim10^{12}~M_\odot$ have higher baryon fractions in the NoAGN simulation because the stellar feedback cannot efficiently push out the gas in massive halos.

\begin{figure}[b!]
\plottwo{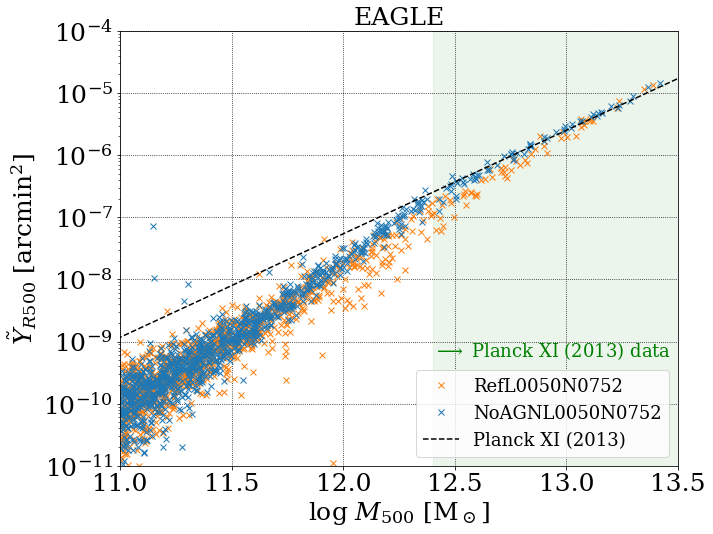}{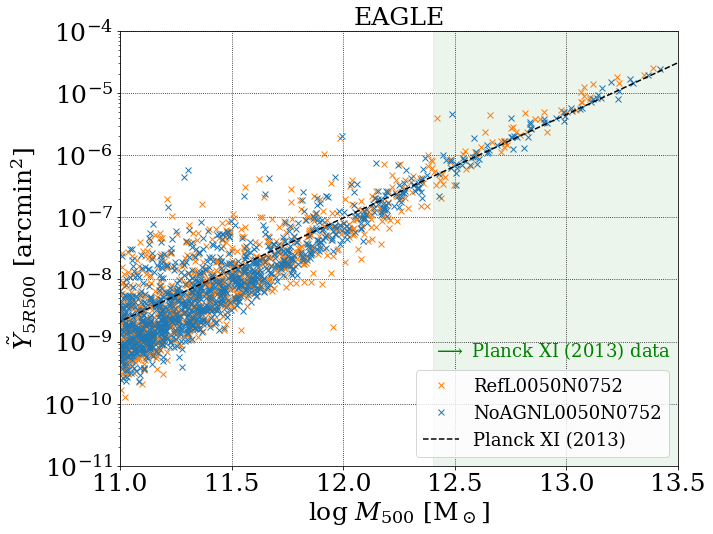}
\caption{The tSZ flux integrated out to $R_{500}$ (left) and $5R_{500}$ (right), for the galaxy samples in EAGLE Ref-L0050N0752 (blue crosses) and NoAGN-L0050N0752 (orange crosses) simulations. The \citetalias{2013A&A...557A..52P} best-fit scaling relation is shown as a dotted line. The shaded green region shows the mass range of the galaxy samples used in the \citetalias{2013A&A...557A..52P} analysis.}
\label{fig:tsz_eagle_refl0050}
\end{figure}

\begin{figure}[t!]
\epsscale{0.5}
\plotone{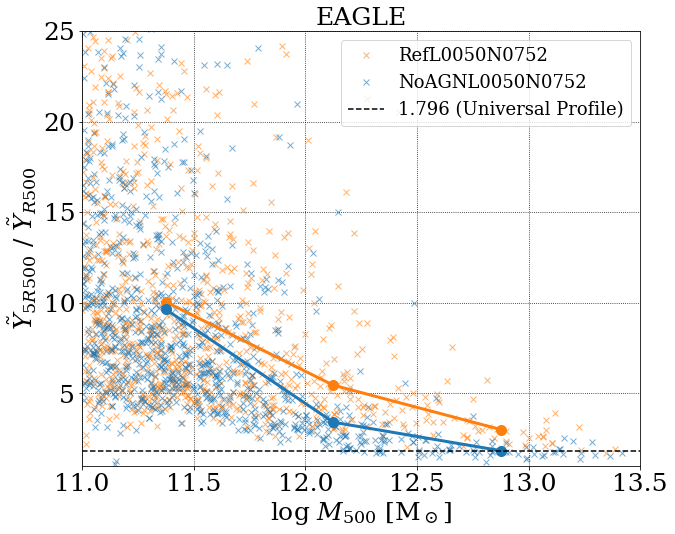}
\caption{$Y_{5R_{500}} / Y_{R_{500}}$ of the EAGLE galaxies in Ref-L0050N0752 and NoAGN-L0050N0752. The dashed line shows the conversion factor 1.796 for the UPP. The blue (Ref-L0050N0752) and orange (NoAGN-L0050N0752) dots connected by the solid lines indicate the median value of the ratios in each $M_{500}$ mass bin. The mass bins are separated by 0.75 dex between $10^{11.0} \textrm{M}_\odot$ and $10^{13.25} \textrm{M}_\odot$.}
\label{fig:tsz_eagle_refl0050_yratio}
\end{figure}

In Figure~\ref{fig:tsz_eagle_refl0050}, we compare the tSZ flux measured within $R_{500}$ and $5R_{500}$. We do not separate the galaxy types here since we are only interested in the behavior depending on the AGN feedback prescription. As in Figure~\ref{fig:tsz_m500}, the Compton-$y$ parameters integrated within the radius $5R_{500}$ better reproduce the scaling relation of \citetalias{2013A&A...557A..52P}. The tSZ fluxes within $R_{500}$ of the reference simulation seem to be lower than in the NoAGN run, in particular in the mid-mass range between $10^{12}$ and $10^{12.5} \textrm{M}_\odot$ (Figure~\ref{fig:tsz_eagle_refl0050}, left panel). This becomes more evident in Figure~\ref{fig:tsz_eagle_refl0050_yratio}, where we show the ratio of the tSZ fluxes within $R_{500}$ and $5R_{500}$. The solid blue and orange lines indicate the median ratio of the galaxy samples within each mass bin, separated by 0.75 dex between $10^{11.0}$ and $10^{13.25} \textrm{M}_\odot$. For both of the simulations, the ratios approach the value from the UPP as the halo mass increases. The median tSZ flux ratios of the reference run are greater than the NoAGN run, indicating that the hot gas has been pushed out beyond $R_{500}$ by the AGN feedback. The median ratios of the NoAGN simulation samples between $10^{12.5}$ and $10^{13.25} \textrm{M}_\odot$ are nearly consistent with the self-similar value. As the halo mass decreases, the stellar feedback becomes efficient in both simulations, and the ratios deviate rapidly from the self-similar value.

\begin{figure}[t!]
    \begin{center}
    \includegraphics[width=.33\textwidth]{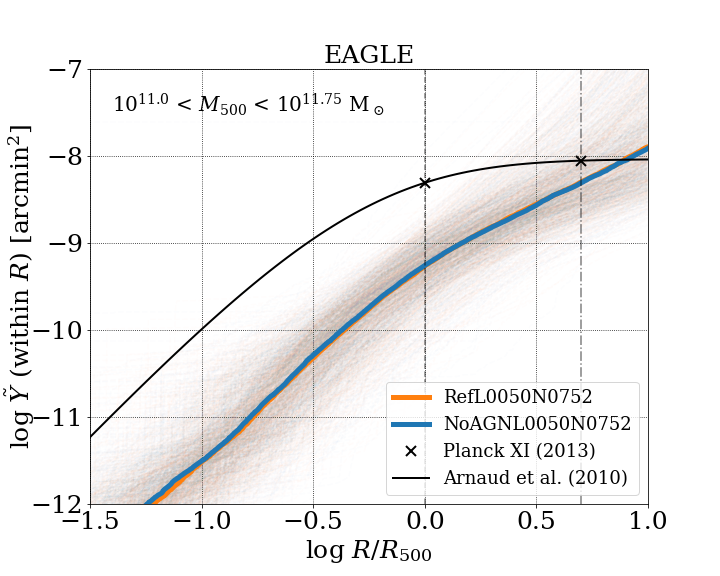}\hfill
    \includegraphics[width=.33\textwidth]{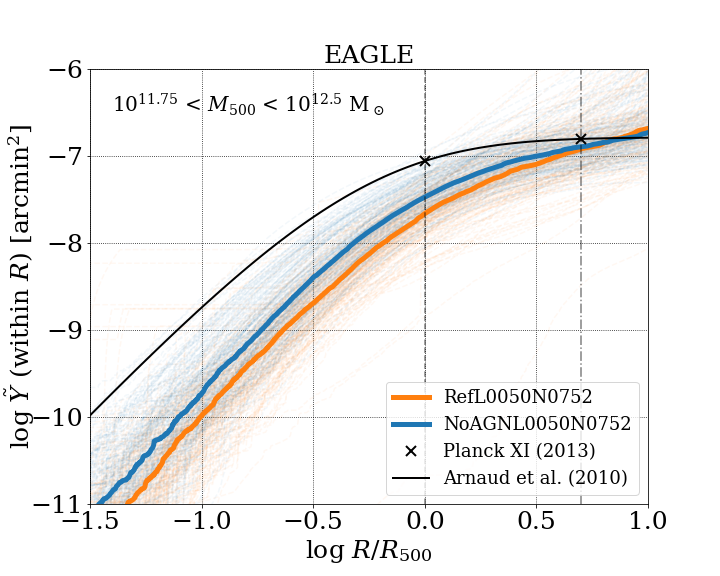}\hfill
    \includegraphics[width=.33\textwidth]{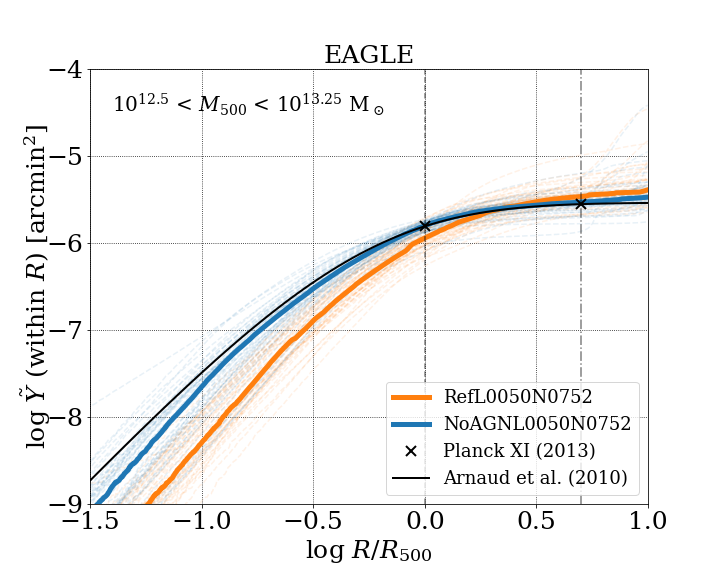}\\
    \includegraphics[width=.33\textwidth]{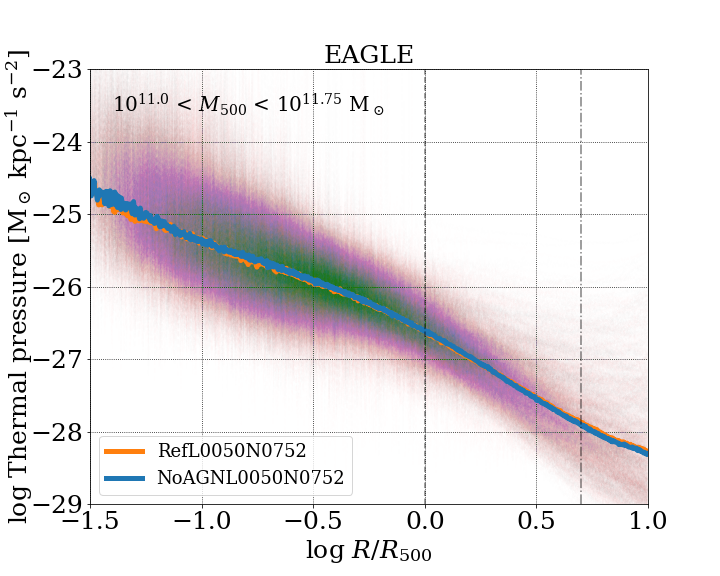}\hfill
    \includegraphics[width=.33\textwidth]{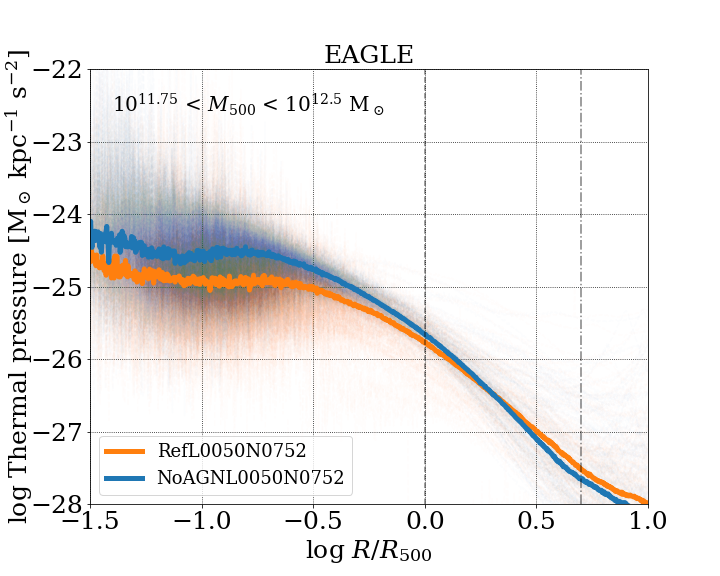}\hfill
    \includegraphics[width=.33\textwidth]{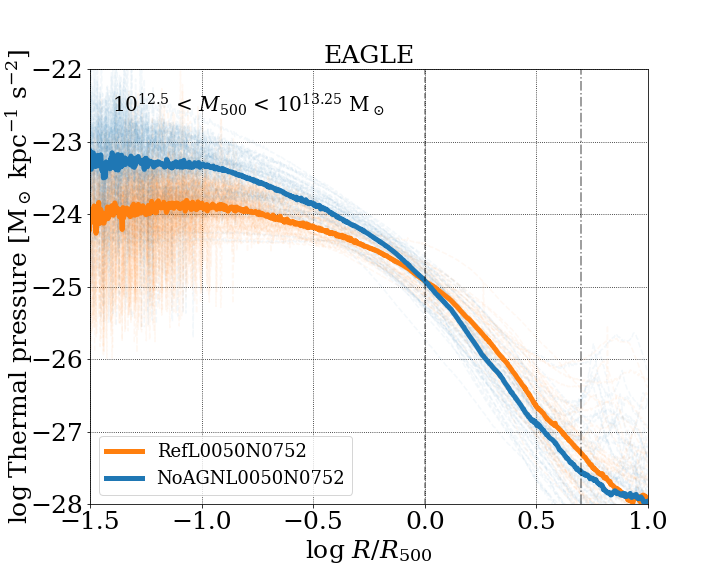}\\
    \includegraphics[width=.33\textwidth]{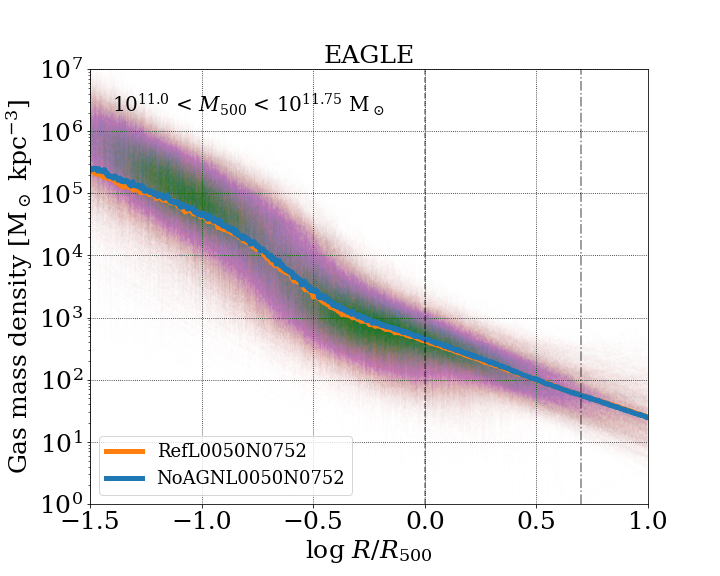}\hfill
    \includegraphics[width=.33\textwidth]{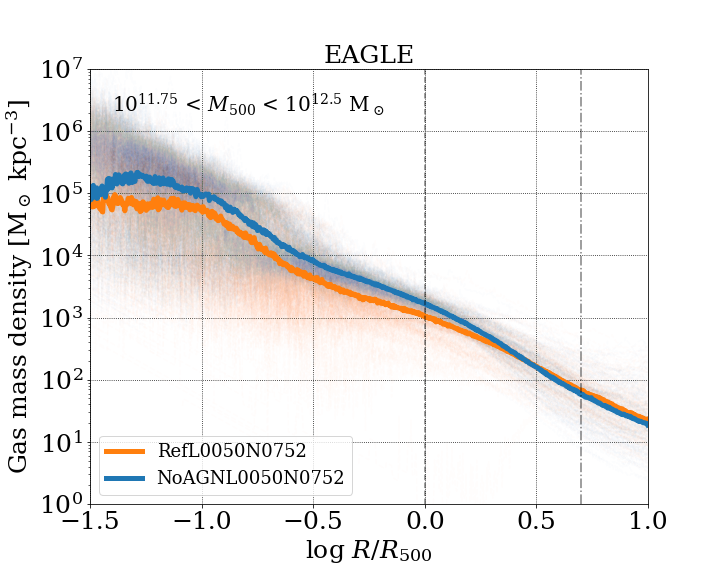}\hfill
    \includegraphics[width=.33\textwidth]{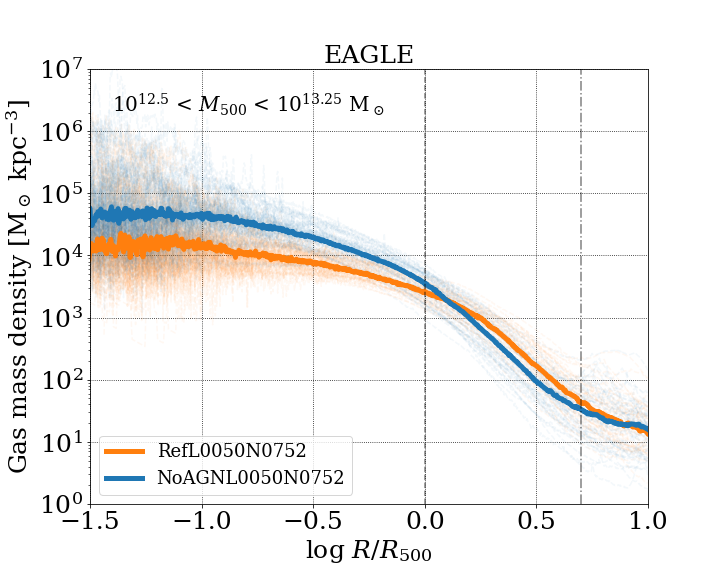}\\
    \includegraphics[width=.33\textwidth]{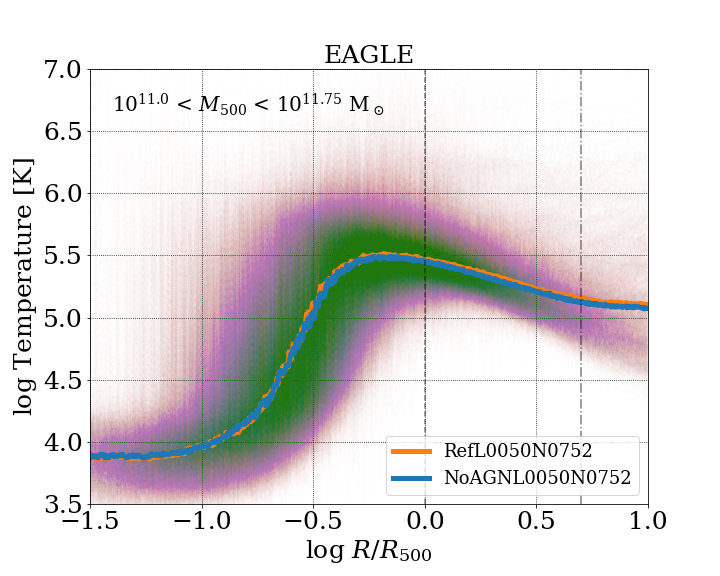}\hfill
    \includegraphics[width=.33\textwidth]{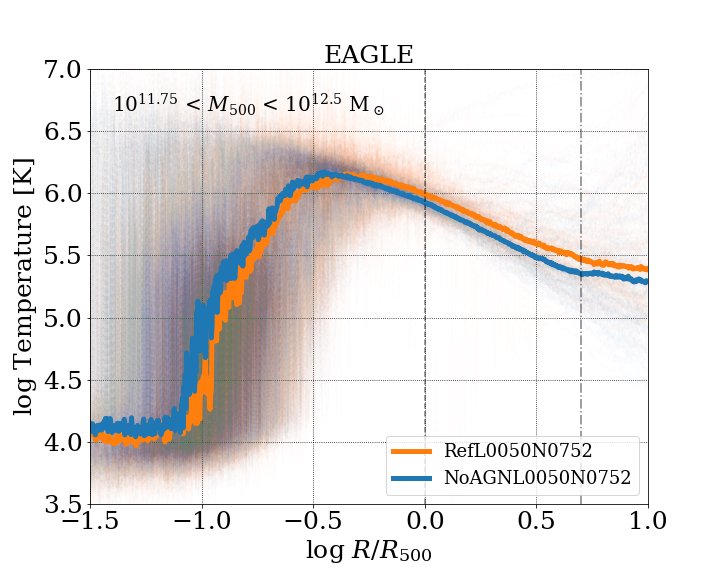}\hfill
    \includegraphics[width=.33\textwidth]{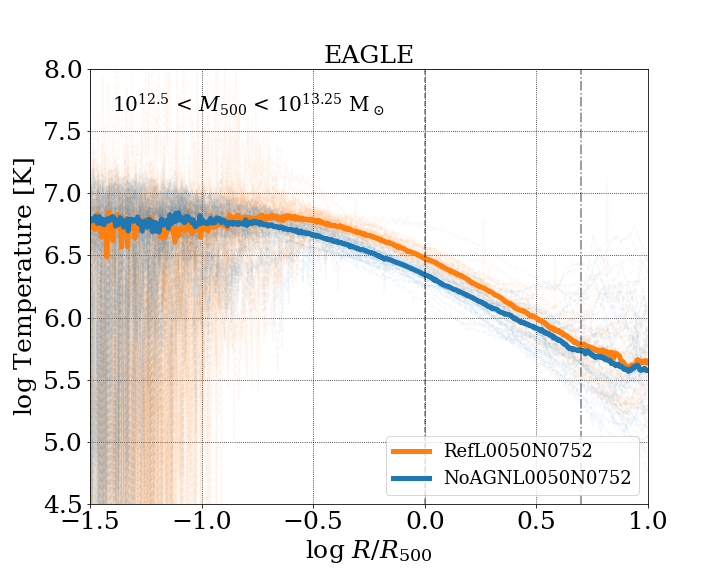}\\
    \caption{Radial profiles of $10^{11.0}\textrm{M}_\odot < M_{500} < 10^{11.75}\textrm{M}_\odot$ (left column), $10^{11.75}\textrm{M}_\odot < M_{500} < 10^{12.5}\textrm{M}_\odot$ (middle column), and $10^{12.5}\textrm{M}_\odot < M_{500} < 10^{13.25}\textrm{M}_\odot$ (right column) galaxies in EAGLE Ref-L0050N0752 (orange) and NoAGN-L0050N0752 (blue) simulations. We show cumulative tSZ flux, thermal pressure, gas mass density, and temperature from top to bottom. The faint dotted lines are the radial profiles of the individual galaxies. The black crosses in the tSZ radial profile plots (top row) denote the tSZ signal at $R_{500}$ and $5 R_{500}$ for $10^{11.375}$, $10^{12.125}$, and $10^{12.875} \textrm{M}_\odot$ mass halos (left to right, respectively), inferred from the self-similar relation of \citetalias{2013A&A...557A..52P}. The mean $R_{500}$ of each mass bin corresponds to 90, 160, 300 physical kpc, respectively. The solid black lines in those plots show spherically integrated tSZ flux as a function of radius, derived from the universal pressure profile \citep{2010A&A...517A..92A}, normalized by the \citetalias{2013A&A...557A..52P} measurement.}
    \label{fig:tsz_eagle_refl0050_radial}
    \end{center}
\end{figure}

In Figure~\ref{fig:tsz_eagle_refl0050_radial}, we present the average radial thermodynamic profiles of the $10^{11.0}\textrm{M}_\odot < M_{500} < 10^{11.75}\textrm{M}_\odot$, $10^{11.75}\textrm{M}_\odot < M_{500} < 10^{12.5}\textrm{M}_\odot$ and $10^{12.5}\textrm{M}_\odot < M_{500} < 10^{13.25}\textrm{M}_\odot$ samples in both of the EAGLE simulations shown above. First of all, we see that the tSZ flux of the two simulations at both $R_{500}$ and $5R_{500}$ is reduced compared to the \citetalias{2013A&A...557A..52P} interpolation as the halo mass decreases, although the flux at $5R_{500}$ better agrees with the UPP profile than at $R_{500}$. As the halo mass increases, the tSZ radial profiles of the Reference and NoAGN runs display a larger discrepancy, especially at $R \lesssim R_{500}$, due to the enhanced AGN feedback. At the highest mass bin, the tSZ profile of the NoAGN simulation is nearly consistent with the UPP profile, implying that gravity is able to overcome feedback and retain the gas close in without AGN feedback. 

Comparing the gas mass density and the temperature profiles in the middle and the right columns, we observe that the reduction in the flux at $R \lesssim R_{500}$ for the Reference simulation comes from the reduced gas mass density (temperature is comparable or even higher), which is a result of the feedback processes pushing the gas out to larger radii. In the lowest mass bin, all the radial profiles of the NoAGN and Reference simulations are comparable because stellar feedback dominates over AGN feedback in this mass range.


\section{Conclusions}\label{sec:conclusion}

We studied the thermodynamic properties of the CGM using the tSZ signal in TNG, EAGLE, and FIRE-2 simulations. We find that the level of agreement of simulations over the mass range $M_{500} = 10^{11}$ to $3 \times 10^{13} \textrm{M}_\odot$ with a self-similar fit inferred from \Planck\ data down to $M_{500} \sim 10^{13} \textrm{M}_\odot$ ($M_\star \sim 10^{11} \textrm{M}_\odot$) is dependent on the aperture size used for the tSZ measurement. We interpret this dependence as evidence that feedback has a significant impact on CGM thermodynamic properties at $R_{500}$ (significant deviations from self-similarity) but that these effects have subsided by $5R_{500}$ (self-similarity is recovered). We also find that the effect is more pronounced in quiescent than in star-forming galaxies, suggesting that the impact of feedback at $R_{500}$ is still in process for star-forming galaxies while it is complete for quiescent ones. 

We find that agreement on radial profiles of gas density and thermal pressure between simulations and ACT analysis at $M_{200} \sim 3 \times 10^{13} \textrm{M}_\odot$ ($z \sim 0.55$) is also radius-dependent.  The disagreement at large radius is consistent with prior suggestions that sub-grid heating in the simulation at large radius is insufficient. Still, it could also be a sign that feedback in the simulations is too effective at these radii.

We also compare the simulated radial profiles of gas density, temperature, and thermal pressure for Milky-way-mass galaxies by separating star-forming and quiescent galaxies. We find support from all three simulations for the above interpretation.  We also find from FIRE-2 simulations that the inclusion of cosmic-rays further enhances these effects, presumably because of the non-thermal pressure supplied by the CR.

Lastly, we compare EAGLE simulations with and without AGN feedback.  We find that the impact of AGN on the $Y\!-\!M_{500}$ scaling relation and on the radial profiles is most significant near $M_{500} \sim 2 \times 10^{12} \textrm{M}_\odot$, vanishing at $10^{11} \textrm{M}_\odot$ and present but smaller above $10^{13} \textrm{M}_\odot$.  At lower masses, the effect is likely due to the decreasing significance of AGN feedback relative to star-formation feedback with decreasing mass, even while star-formation feedback remains sufficient to cause a deviation from self-similarity. At higher masses, gravity has greater ability to counter the effect of feedback.

\section{Future Work}\label{sec:futurework}

We will extend this work to the simulation study of the CGM using other multi-wavelength probes. In addition to the tSZ effect measurement, X-ray observation is another tool to characterize the hot gas properties around the galaxy groups and clusters \citep[e.g.,][]{2015MNRAS.449.3806A, 2016MNRAS.463.4533V, 2020AN....341..177L}. Similar to the $Y\!-\!M$ relation, the soft X-ray luminosity of the galaxy clusters are expected to follow the mass scaling relation. Several observations suggest that the emission is affected by the non-gravitational heating, such as the AGN feedback \citep[e.g.,][]{2009A&A...498..361P, 2011A&A...536A..10P, 2015MNRAS.449.3806A}. Using the soft X-ray luminosity allows us to study the mass-dependent heating mechanism around the galaxies.

Besides the tSZ and X-ray imaging, several observational tools have been emerging to explore the hot halo across the electromagnetic spectrum. They include X-ray spectroscopy, which can probe the gas metallicity and thus the chemical composition of the CGM \citep{2021arXiv210211510V}, and fast radio bursts (FRBs), which provide the line-of-sight free electron density with its frequency-dependent dispersion measure \citep[e.g.,][]{2019ApJ...872...88R, 2019MNRAS.485..648P, 2020Natur.581..391M, 2021arXiv210713692C}. Traditional UV absorption line spectroscopy can also trace the hot gas via high-energy ions such as O VI and Ne VIII. However, none of these methods can individually link the corresponding observation to a complete set of physical properties of the CGM, such as density, temperature, and metallicity. The best opportunity to gather the spatial and thermodynamic information of the hot CGM will be available through combining the observations across multiple wavelengths. The comprehensive approach will allow us to compare the relative importance of the feedback mechanisms that affect the regulation of star formation at various lengths and mass scales. In future work, we will address how the CGM can be used to distinguish different feedback models and how the relative importance of those models changes with the physical properties of galaxies. We will also study what combinations of the observational probes give the best understanding of the gas properties to break the degeneracy among them.

\software{\texttt{astropy} \citep{2013A&A...558A..33A}, \texttt{numpy} \citep{2020NumPy-Array}, \texttt{matplotlib} \citep{2007CSE.....9...90H}, \texttt{Mop-c GT} (``Model-to-observable projection code for Galaxy Thermodynamics'')\footnote{https://github.com/samodeo/Mop-c-GT}, \texttt{scipy} \citep{2020SciPy-NMeth}}\\

\begin{acknowledgements}
We thank Lee Armus for useful discussion and carefully reading the manuscript. JK is supported by a Robert A. Millikan Fellowship from California Institute of Technology (Caltech). JK, JGB, SG, NB, and JCH acknowledge support from the Research and Technology Development fund at the Jet Propulsion Laboratory through the project entitled “Mapping the Baryonic Majority”. NB, EM, and SA acknowledges support from NSF grant AST-1910021. Support for PFH was provided by NSF Research Grants 1911233 \&\ 20009234, NSF CAREER grant 1455342, NASA grants 80NSSC18K0562, HST-AR-15800.001-A. Numerical calculations were run on the Caltech compute cluster ``Wheeler,'' allocations FTA-Hopkins/AST20016 supported by the NSF and TACC, and NASA HEC SMD-16-7592. JCH acknowledges support from NSF grant AST-2108536. We acknowledge the Virgo Consortium for making their simulation data available. The EAGLE simulations were performed using the DiRAC-2 facility at Durham, managed by the ICC, and the PRACE facility Curie based in France at TGCC, CEA, Bruy\`{e}resle-Ch\^{a}tel. A portion of the research described in this paper was carried out at the Jet Propulsion Laboratory, California Institute of Technology, under a contract with the National Aeronautics and Space Administration (NASA). This research has made use of NASA's Astrophysics Data System Bibliographic Services.
\end{acknowledgements}

\newpage
\appendix

\renewcommand{\thefigure}{A\arabic{figure}}
\renewcommand{\theHfigure}{A\arabic{figure}}
\setcounter{figure}{0}
\restartappendixnumbering

\section{Computing tSZ signal using particle data of cosmological simulations}\label{sec:appendix_tszcalc}

Using particle data of the cosmological simulations, we calculate a total Comptonization parameter within a certain radius by computing each particle's tSZ signal strength. First, we calculate the electron number density
\begin{equation}
    n_e = \frac{\rho_\textrm{gas}}{\mu_e m_\textrm{H}}.
\end{equation}
The tSZ signal of a particular gas cell, $\Upsilon$, is
\begin{equation}
    \begin{split}
        \Upsilon &= n_e \sigma_T \frac{k_B T_e}{m_e c^2}\\
        &= \left(\frac{m_\textrm{gas}}{\mu_e m_\textrm{H}} \right) \sigma_T \frac{k_B T_e}{m_e c^2},
    \end{split}
\end{equation}
where $\mu_e$ is the mean molecular weight per free electron in units of the hydrogen mass (i.e., amu). Then, we integrate the signals for all the gas particles within the radius, for example, $R_{500}$,
\begin{equation}
    Y_{R_{500}} {D^2_A (z)} = \sum \Upsilon (r < R_{500}).
\end{equation}

The TNG dataset gives the internal energy of each gas cell and we need to convert it to a gas temperature. The mean molecular weight $\mu$ there is defined as the average mass per particle:
\begin{equation}
    \mu = \frac{\left(\sum_{j} n_j m_j\right) + n_e m_e}{\left(\sum_{j}n_j\right) + n_e} \frac{1}{m_p},
\end{equation}
where $n_j$, $m_j$ are the number density of atoms and the corresponding mass, respectively. If we consider hydrogen ($n_\textrm{H}, m_p$), helium ($n_\textrm{He}, 4 m_p$), and all the other metals ($n_z, m_z$) with an approximation $m_z \sim 4 m_p$, the total gas mass density $\rho_\textrm{gas} = n_\textrm{H} m_p + n_\textrm{He} (4 m_p) + n_z m_z$ and
\begin{equation}
    \begin{split}
        \mu &= \frac{n_\textrm{H} m_p + n_\textrm{He} (4 m_p) + n_z m_z}{n_\textrm{H} + n_\textrm{He} + n_z + n_e} \frac{1}{m_p} \\
        &= \frac{1}{\left[n_\textrm{H} + n_\textrm{He} + \left(\frac{\rho_\textrm{gas}}{m_z} - \frac{1}{4}n_\textrm{H} - n_\textrm{He} \right) + n_e\right]/\rho_\textrm{gas}} \frac{1}{m_p}\\
        &= \frac{4}{1 + 3 X + 4 X x_e},
    \end{split}
\end{equation}
where $x_e = n_e / n_\textrm{H}$ is the electron abundance. $X = (n_\textrm{H} m_p)/\rho_\textrm{gas} = 0.76$. (The EAGLE simulation does not provide electron abundance, so we assume fully ionized gas following the approach of \citealt{2018MNRAS.480.4017L} for the simulation.) Then, the (specific) internal energy of the cell
\begin{equation}
    u = \frac{1}{\gamma-1} \left(\frac{1}{\mu}\right) k_B T_e,
\end{equation}
where the adiabatic index $\gamma = 5/3$, and the gas temperature of the cell is
\begin{equation}
    T_e = (\gamma - 1) \frac{u}{k_B} \mu.
\end{equation}

In Section~\ref{subsec:radial}, where we compare the radial tSZ profiles of $\sim 10^{12}\ \textrm{M}_\odot$ halos, we present the thermal pressure profiles as well. Note that the electron pressure $P_e$ and the thermal pressure $P_\textrm{th}$ are related by
\begin{equation}
    P_e \simeq \left(\frac{2 + 2 X}{3+5 X}\right) P_\textrm{th}
\end{equation}
\citep{2017JCAP...11..040B}, and we can directly calculate the thermal pressure from the particle data using the gas mass, internal energy per unit mass, and the volume of the gas cell.

\section{Effect of neighboring particles in calculating radial profiles}\label{sec:appendix_twohalo}

In the reconstruction of radial profiles and $Y$ vs. $R$ in this paper from TNG and EAGLE, we do not limit ourselves to the particles in the halo or subhalo being considered but rather include all particles; that is, neighboring halos, subhalos, and the particles not bound to any halo/subhalo are included. This appendix justifies the choice. The large-volume TNG and EAGLE simulation datasets provide two kinds of group catalogs. Those are halo (group) and subhalo (galaxy) catalogs derived by the FoF \citep{1982ApJ...257..423H, 1985ApJ...292..371D} and \texttt{subfind} algorithms \citep{2001MNRAS.328..726S}, respectively. Simulation studies often limit the analysis to the particles associated with a single halo or subhalo. However, as shown in Figure~\ref{fig:eagle_act2020tsz}, the effect of the `two-halo' term becomes not negligible at large radii from the center of the galaxy \citep{2017MNRAS.467.2315V, 2018PhRvD..97h3501H, 2021ApJ...919....2M}.

\begin{figure}[b!]
\plottwo{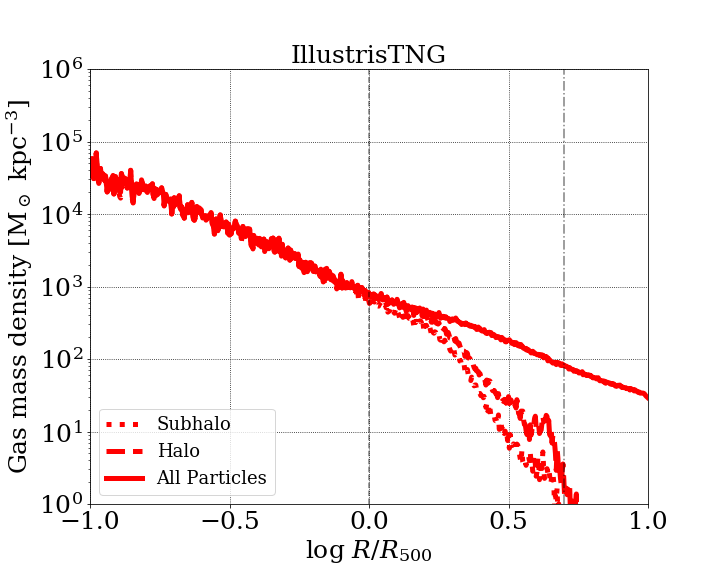}{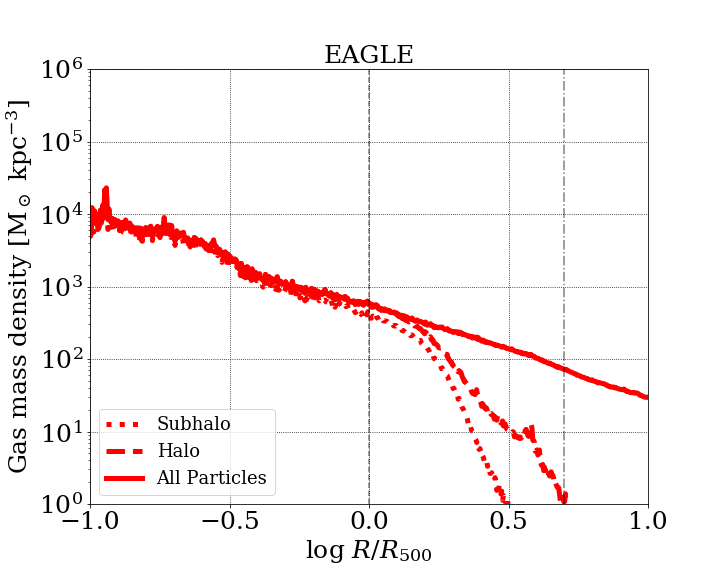}\\
\plottwo{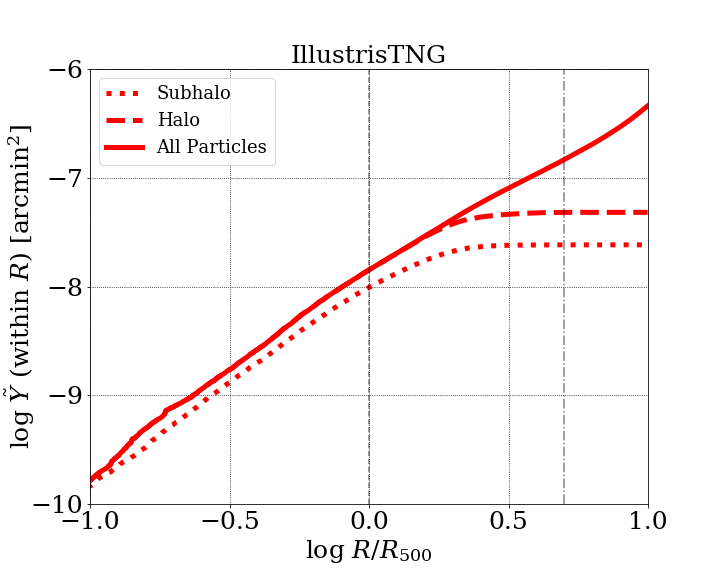}{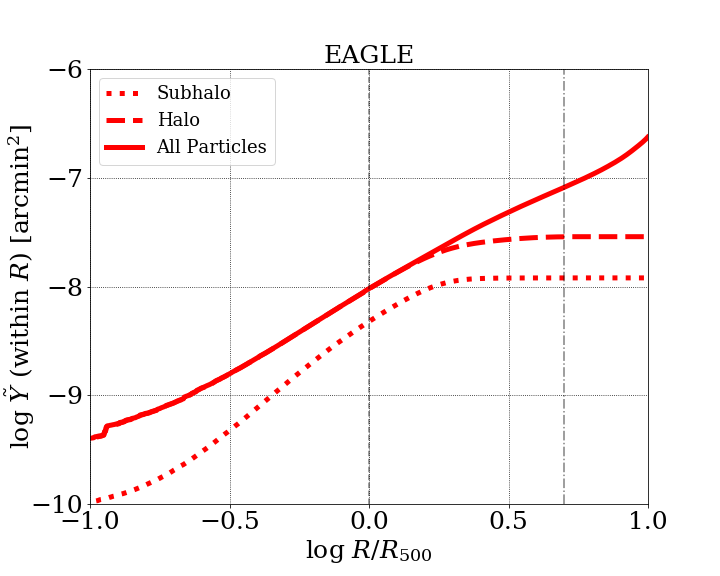}
\caption{Average radial gas mass density (upper) and tSZ flux (lower) of the $\sim 10^{12} \textrm{M}_\odot$ galaxies in TNG and EAGLE. Dotted, dashed, and solid lines show the radial profiles calculated using particles within a subhalo, a halo group, and the entire particles within the simulation volume around the center of the galaxy, respectively. The figures indicate that the discrepancy in the radial profiles exists depending on the particle selection in both simulations. It implies that the inclusion of the two-halo term is essential in comparing the simulation models to the observations, particularly at large radii from the center. Note that we show the profiles to contrast the particle selections in deriving the radial profiles. Readers need to avoid the physical interpretation comparing the radial profiles of the simulations.}
\label{fig:appendix_twohalo}
\end{figure}

Figure~\ref{fig:appendix_twohalo} shows the average radial gas mass density and tSZ flux profiles in TNG (TNG100-3) and EAGLE (RefL0050N0752) simulations. We randomly sampled $\sim$ 50 MW-sized, quiescent galaxies in each simulation and calculated the tSZ flux and gas mass density profiles similar to Figure~\ref{fig:mw_radial}. The dotted, dashed, and solid lines in the Figure show the average radial profiles calculated using particle data affiliated to a single (central) subhalo, the parent FoF halo, and the entire particle data in the simulation volume, respectively. As illustrated in Figure~2 of \cite{2021ApJ...919....2M}, where they showed the gas mass density and thermal pressure profiles, the two-halo term starts to contribute to the radial profile beyond $R_{200}\sim 1.5R_{500}$. It is clear that the $Y$ vs. $R$ and radial profiles would show a striking and nonintuitive slope change beyond $R_{500}$ where particles from neighboring halos are not included. It would be impossible to reproduce the \citetalias{2013A&A...557A..52P} results without including this effect, which is sensible since on-sky SZ measurements cannot distinguish between one halo and its neighbors. Even at $R_{500}$, one begins to see a distinction in the radial profiles, and to a lesser extent in $Y$ vs. $R$, of considering only the central subhalo rather than the larger halo in which it resides. The figures make it clear that consideration of the CGM outskirts requires the inclusion of neighboring halos (i.e., all particles).

\bibliography{ms}{}
\bibliographystyle{aasjournal}

\end{document}